\documentclass[twocolumn]{aastex631}

\DeclareUnicodeCharacter{2212}{-}

\shorttitle{Lithium Abundances in M48}
\shortauthors{Sun et al.}

\graphicspath{{./}{figures/}}

\begin{document}

\title{WIYN$^1$ Open Cluster Study$^2$ 89. M48 (NGC 2548) 2: Lithium Abundances in the 420 Myr Open Cluster M48 From Giants Through K Dwarfs}

\footnote[1]{The WIYN Observatory is a joint facility of the University of Wisconsin Madison, Indiana University, the National Optical Astronomy Observatory and the University of Missouri.}
\footnote[2]{WIYN Open Cluster Study, \citet{2000ASPC..198..517M}.}

\correspondingauthor{Qinghui Sun}
\email{qingsun@tsinghua.edu.cn}

\author[0000-0003-3281-6461]{Qinghui Sun}
\affiliation{Department of Astronomy, Tsinghua University, Beijing, 100084, China}
\affiliation{Department of Astronomy, Indiana University, Bloomington, IN 47405, USA}

\author[0000-0002-3854-050X]{Constantine P. Deliyannis}
\affiliation{Department of Astronomy, Indiana University, Bloomington, IN 47405, USA}

\author[0000-0002-5719-5596]{Aaron Steinhauer}
\affiliation{Department of Physics and Astronomy, State University of New York, Geneso, NY 14454, USA}

\author[0000-0001-8841-3579]{Barbara J. Anthony-Twarog}
\author[0000-0001-5436-5206]{Bruce A. Twarog}
\affiliation{Department of Physics and Astronomy, University of Kansas, Lawrence, KS 660045, USA}

\begin{abstract}

We consider WIYN/Hydra spectra of 329 photometric candidate members of the 420-Myr-old open cluster M48, and report Lithium detections or upper limits for 234 members and likely members. The 171 single members define a number of notable Li-mass trends, some delineated even more clearly than in Hyades/Praesepe: The giants are consistent with subgiant Li dilution and prior MS Li depletion due to rotational mixing. A dwarfs (8600-7700K) have upper limits higher than the presumed initial cluster Li abundance. Two of five late A dwarfs (7700- 7200K) are Li-rich, possibly due to diffusion, planetesimal accretion, and/or engulfment of hydrogen-poor planets.  Early F dwarfs already show evidence of Li depletion seen in older clusters. The Li-$T_{\rm eff}$ trends of the Li Dip (6675-6200K), Li Plateau (6200-6000K), and G and K dwarfs (6000-4000K) are very clearly delineated and are intermediate to those of the 120-Myr-old Pleiades and 650-Myr-old Hyades/Praesepe, which suggests a sequence of Li depletion with age. The cool side of the Li Dip is especially well-defined with little scatter. The Li-$T_{\rm eff}$ trend is very tight in the Li Plateau and early G dwarfs, but scatter increases gradually for cooler dwarfs. These patterns support and constrain models of the universally dominant Li depletion mechanism for FGK dwarfs, namely rotational mixing due to angular momentum loss; we discuss how diffusion and gravity-wave driven mixing may also play roles. For late-G/K dwarfs, faster rotators show higher Li than slower rotators, and we discuss possible connections between angular momentum loss and Li depletion.

\end{abstract}

\section{Introduction} \label{sec:intro}

Inside stars, Lithium (Li) is destroyed by energetic protons at temperatures exceeding 2.5 $\times$ 10$^6$ K, and thus survives only in the outermost layers. The surface Li abundances, A(Li)\footnote[3]{A(Li) = 12 + log(N$_{Li}$/N$_H$), where N$_X$ is number of atoms of species X.}, thereby provide a direct observational tool to study physical processes occurring below.  {\it Any} process affecting Li at the base of the Surface Convection Zone (SCZ) also affects the measured A(Li).  It is useful to use Standard Stellar Evolution Theory (SSET\footnote[4]{No rotation, magnetic fields, mass loss or gain, and here, diffusion.}) as a reference (\citealt{1990ApJS...73...21D}).  In SSET, A(Li) can be affected by the nuclear destruction of Li at the base of the SCZ, and by subgiant dilution as the SCZ deepens past the boundary of the Li preservation region.  For masses $\gtrapprox$ $0.8M_{\odot}$, destruction occurs during the early pre-main sequence (pre-MS) only, whereas for lower masses destruction can extend to the MS. The depth of the SCZ increases as stellar mass decreases as does duration of the Li burning phase; as a result, surface Li depletion is minimal for F dwarfs and earlier types, more substantial for G dwarfs, and quite dramatic for K dwarfs and later types.  Standard theory also predicts greater Li depletion for greater metallicity (at a given mass).  Finally, in SSET, there should be no differences in A(Li) in dwarfs of the same age, composition, and mass (equivalently $T_{\rm eff}$).

In sharp contradistinction to the remarkable agreement between the standard solar model (including some helium diffusion) and helioseismology (\citealt{1995RvMP...67..781B, 2022A&A...661A.140M}), the solar $A(Li) \sim 1.05$ (\citealt{1997AJ....113.1871K}) lies {\it a factor of 50} below SSET-predicted depletion of a factor of 3 (\citealt[hereafter P97]{1997ARA&A..35..557P}).  Even worse, the vast majority of dwarf Li abundances lie below standard predictions (\citealt[][hereafter C17]{2017AJ....153..128C}). Clearly, Li abundances reveal the action of physical mechanisms not included in SSET.

Star clusters are ideal tools to help identify which physical mechanisms are at work because they are coeval populations with determinable ages where stars have the same initial composition in a given cluster.  The lower envelope of the relatively simple Li-$T_{\rm eff}$ relation in the 120 Myr-old Pleiades can be matched by SSET (\citealt{2014ApJ...790...72S}); however, even in the Pleiades, SSET fails to account for spreads in A(Li) at a given $T_{\rm eff}$, and the large Li overabundances seen in rapidly rotating late G and K dwarfs. Older clusters such as the 650 Myr-old Hyades and Praesepe reveal additional Li-$T_{\rm eff}$ features that form during the MS, none of which can be explained by SSET. These include (Figure \ref{fig:ALi_teff}): a) Li depletion in late-A stars, b) severe Li depletion in F dwarfs (the Li Dip; \citealt{1986ApJ...302L..49B}), c) modest Li depletion in late F/early G stars (the Li Plateau, C17), and striking Li depletion in G and K dwarfs (\citealt{1997MNRAS.292..177J, 2002MNRAS.336.1109J}).  Importantly, stars of all these spectral types spin down during the MS, so Li depletion is correlated with spin-down. We stress that, contrary to long-held beliefs that stars hotter than the break in the Kraft curve do not spin down (\citealt{1970saac.book..385K}), late-A dwarfs do, in fact, spin down though perhaps on a longer time scale than cooler dwarfs (\citealt{2019AJ....158..163D}).

Aided also by observations of Beryllium (Be) and Boron (B), which survive to deeper layers (temperatures of 3.5 $\times$ 10$^6$ K and 5 $\times$ 10$^6$ K, respectively), the preponderance of evidence favors rotational mixing induced by stellar spin-down as the dominant Li depletion mechanism across all these spectral types. Besides the correlation between spin-down and Li depletion mentioned above, some of this evidence includes the following. For the Li Dip evidence includes its early formation (\citealt{2004ApJ...614L..65S}), the Li/Be depletion correlation (\citealt{1998ApJ...498L.147D, 2001ApJ...553..754B, 2004ApJ...613.1202B, 2022ApJ...927..118B}), the Be/B depletion correlation (\citealt{1998ApJ...492..727B, 2005ApJ...621..991B, 2016ApJ...830...49B}), and revelation of the internal Li and Be profiles in subgiants evolving out of the Li Dip (\citealt{2000ApJ...544..944S, 2020ApJ...888...28B}). For the Li Plateau and G dwarfs evidence includes the MS Li depletion (C17) and higher A(Li) observed in Short Period Tidally Locked Binaries (\citealt{1990PhDT.........3D, 1994ApJ...434L..71D, 1995ApJ...453..819R}).  For rapidly rotating, young late G/K dwarfs, effects of magnetic fields (\citealt{2013ApJ...765..126M, 2014ApJ...789...53F}) in inflating stellar radii (\citealt{2018MNRAS.476.3245J, 2019MNRAS.483.1125J, 2021MNRAS.500.1158J}) together with effects of rapid rotation may need to be taken into account (\citealt{2015MNRAS.449.4131S, 2015ApJ...807..174S}). Other proposed mechanisms include diffusion, mass loss/gain, and mixing by gravity waves, and some of these may at times play a role (C17).

With an age roughly halfway between that of the Pleiades and that of the Hyades and Praesepe, M48 (420 $\pm$ 30 Myr; Deliyannis et al. 2023, in preparation) provides an excellent opportunity to study the development of {\it all} the various features in the Li-$T_{\rm eff}$ relation across all masses from the turnoff through K dwarfs, and for the few cluster giants. Our study of Li in M48 will thus further test proposed mechanisms and guide future models. To help delineate the properties of the Li-$T_{\rm eff}$ trend more precisely and to increase the probability of finding effects that might be relatively rare, we have observed a large sample of 329 candidate member stars.

\section{Observations and Data Reductions}  \label{sec:obs}

In M48 Paper 1 (\citealt{2020AJ....159..220S}), we reported radial ($V_{\rm RAD}$) and rotational ($v$ sin $i$) velocities for 287 photometrically selected candidate members observed with WIYN/Hydra. Using our spectra and Gaia DR2 (\citealt{2018A&A...616A...1G, https://doi.org/10.26131/irsa12}) proper motion and parallax information, we evaluated multiplicity and membership for each.  For multiplicity we defined three designations: ``s" -- probable single star, ``b" -- probable binary (or multiple), and ``u" -- uncertain multiplicity (this was designated as ``?" in Paper 1).  For membership we defined ``m" -- member, ``lm" -- likely member (``m?" in Paper 1), ``ln" -- likely non-member, and ``n" -- non-member (``n?" in Paper 1). Combining these designations for the 287 stars led to 152 single members (sm), 11 binary members (bm), 16 members of uncertain multiplicity (um), 56 single non-members (sn), 28 single likely non-members (sln), 2 single likely members (slm), 1 binary likely member (blm), 5 binary non-members (bn), 10 likely members of uncertain multiplicity (ulm), 3 non-members of uncertain multiplicity (un), and 3 likely non-members of uncertain multiplicity (uln). A detailed description of the designations can also be found in Table \ref{tab:tally}.  We reported [Fe/H] for slowly rotating sm stars with $\sigma$($T_{\rm eff}$) $< 75K$ from $T_{\rm eff}$ derived using multiple photometric indices, and found no dependence on $T_{\rm eff}$ over a range of 2500 K in $T_{\rm eff}$.  We also reported a cluster average [Fe/H] = -0.063 $\pm$ 0.007 dex ($\sigma_{\mu}$, and $\sigma$ = 0.151 dex).

In this study we report Li abundances for stars from Paper 1. We have enlarged the sample with 42 Gaia candidate proper motion and parallax members that were not previously observed (see Figure 5 of Paper 1); 28 stars were observed again, increasing their signal-to-noise (SNR). Table \ref{tab:config} describes the four new WIYN/Hydra configurations, which were observed using the same instrument settings, and Table \ref{tab:obs} shows the observing logs. The data were processed and reduced in the same way as Paper 1.
\vspace{-5 mm}
\begin{deluxetable}{ccccc}
	\label{tab:config}
	\tablecaption{Hydra configurations}
	\tabletypesize{\tiny}
	\tablehead{
		\colhead{Description} &
		\colhead{Configuration} &
		\colhead{$V$ range} &
		\colhead{$B-V$ range} &
		\colhead{\# of Stars$^1$} \\
		\colhead{} &
		\colhead{name} &
		\colhead{mag} &
		\colhead{mag} &
		\colhead{ }
	} 
	\startdata
	\hline
	bright & m48gb & 9.477- 13.631 & 0.076 - 0.528 & 13 \\ 
	faint 1 & m48gf1 & 12.082 - 16.090 & 0.274 - 1.014 & 43 \\
	faint 2 & m48gf2 & 12.358 - 16.021 & 0.400 - 0.998 & 13 \\
	faint 3 & m48gf3 & 12.812- 16.090 & 0.426 - 1.014 & 18 \\
	\hline
	\enddata
	\tablecomments{1. Configurations m48gf2 \& m48gf3 both included star 2119, 3016, 3021, 3025, 3030, 3033, 3041, 3044, 3058, 3071; configurations m48gf1 \& m48gf3 both included star 3020, 3049, 3061, 3069, 3070, 3073; configurations m48gb \& m48gf3 both included star 3034.
	2. Star 2014, 2019, 2038, 2049, 2060, 2061, 2076, 2080, 2091, 2095, 2115, 2119, 2125, 2126, 2130, 2135, 2140, 2141, 2148, 2165, 2178, 2203, 2232, 2235, 2238, 2267, 2274, 2281 have already been observed in Paper 1, they were observed again in these configurations.}
\vspace{-10mm}
\end{deluxetable}

\begin{deluxetable*}{lllrrrrrrlll}
	\label{tab:obs}
	\tablecaption{M48 observing logs}
	\tabletypesize{\scriptsize}
	\tablehead{
		\colhead{Date$^1$} &
		\colhead{Configurations} &
		\colhead{Exposure Time$^2$} &
		\colhead{Standards$^3$} &
		\colhead{$ < 1\sigma\ ^4$} &
		\colhead{$1 - 2\sigma$} & 
		\colhead{$ > 2\sigma$} & 
		\colhead{$V_{\rm{RAD}}$ (km s$^{-1}$) $^5$} &
		\colhead{$\sigma_{V_{\rm{RAD}}}$ (km s$^{-1}$) $^5$}
	} 
	\startdata
	\hline
	16 Jan 2020 & m48gb & 4.35 hr & yes & 1 & 0 & 0 & 8.90 & 0.90 \\ 
	17 Jan 2020 & m48gb, m48gf1 & 6.3 hr, 1 hr & yes &  2& 0 & 0 & 8.14, 8.02 & 1.01, 1.10 \\
	18 Jan 2020 & m48gf1 & 8.47 hr & yes & 1 & 1 &  0& 8.42 & 0.66 \\ 
	23 Jan 2020 & m48gf2 & 6.17 hr & no & 0 & 0 & 0 & 8.60 & 1.07 \\ 
	13 Feb 2020 & m48gf3 & 6.5 hr & yes & 5 & 0 & 0 & 8.08 & 1.14 \\
	\hline
	\enddata
	\tablecomments{1. UT date when afternoon calibrations began. 
		2. The total exposure for a given configuration.
		3. Whether radial velocity standard was observed in that night.
		4. The number of radial velocity standards that fall within $1\sigma$, between $1\sigma$ and $2\sigma$, and above $2\sigma$ compared to the literature (Gaia $V_{\rm RAD}$, \citealt{2018AA...616A...7S}).
		5. Average radial velocity and standard deviation of each configuration. We calculate arithmetic mean and standard deviation for configurations m48gb, m48gf2, and m48gf3 after eliminating outliers. We fit a Gaussian profile to m48gf1 stars, the average and 1$\sigma$ error for m48gf1 is from the Gaussian fit.}
	    \vspace{-10mm}
\end{deluxetable*}

\section{Velocities and stellar parameters} \label{sec:parameter}

In this section we follow the methods of Paper 1 to evaluate final multiplicity and membership of the newly observed stars.  We then discuss the adopted stellar parameters.

\subsection{Radial and Rotational Velocities, Multiplicity, and Membership}   \label{sec:rad}

For each configuration, the co-added spectra were run through the {\it fxcor} task in IRAF\footnote[5]{IRAF is distributed by the National Optical Astronomy Observatories, which are operated by the Association of Universities for Research in Astronomy Inc., under cooperative agreement with the National Science Foundation.} to derive radial and rotational velocities. After eliminating outliers, we computed the average $V_{\rm RAD}$ for that configuration in the given night, except for m48gf1, where we fit a Gaussian profile to the histogram to derive average $V_{\rm RAD}$ and 1$\sigma$ error. We then shifted the night's average wavelength to match the cluster's average $V_{\rm RAD}$ of 8.53 $\pm$ 0.05 km s$^{-1}$ (from Paper 1), and combined the shifted spectra of the same configuration from different nights. Table \ref{tab:new_parameter} shows the final $V_{\rm RAD}$ and {\it v} sin {\it i}. For stars observed on multiple nights, we report $V_{\rm RAD}$ (and $\sigma$) for all individual nights (Columns 7 and 8). By cross-correlating the star's spectrum with a template, {\it fxcor} computes $V_{\rm RAD}$ for all the spectral lines in the object spectrum, and then reports the mean $V_{\rm RAD}$ and $\sigma$ by fitting a Gaussian profile to the $V_{\rm RAD}$ distribution in Fourier space. We compare $V_{\rm RAD}$ measurements for the same star from different nights, and examine the power spectrum from {\it fxcor}. If the $V_{\rm RAD}$s from different nights disagree by more than 2$\sigma$ (using the largest individual error), or if the {\it fxcor} power spectrum shows two (or multiple) peaks, we assigned the star as a binary. Thus, multiplicity (and membership) were evaluated in the same way as Paper 1.  Multiplicity and membership (mm) status is indicated in Column 13. Our rotational velocities ({\it v} sin {\it i}) and errors are calculated based on line broadening from {\it fxcor}, shown in column 11 and 12. Please see more extended discussions about $V_{\rm RAD}$, {\it v} sin {\it i}, and multiplicity from our M48 data in Section 3 of Paper 1, and also the broader discussion about cluster binary fractions.

Column 9 of Table 5 shows the periods of \citet[hereafter B15]{2015AA...583A..73B} for stars in common. Column 11 shows the periods derived by using the TESS light curves (\citealt{https://doi.org/10.26134/exofop3}), which are retrieved through the Mikulski Archive for Space Telescopes (MAST) portal\footnote[6]{\url{https://mast.stsci.edu/portal/Mashup/Clients/Mast/Portal.html)}}. We then follow the same procedure as described in \citet{2022RAA....22g5008S} to compute the autocorrelation functions to derive rotational periods. For those stars with both B15 and TESS periods, the vast majority are in excellent agreement.  The stars that disagree all tend systematically toward shorter TESS period, which could potentially introduce undesirable systematic errors in the analysis.  However, all such stars have $\sigma_{TESS} \geq$ 1.15 d and/or $\sigma_{B15} \geq$ 0.75 d, so in Section \ref{sec:discuss} we consider only periods with smaller errors. A comparison of our {\it v} sin {\it i} using equatorial velocities from these periods, with radii were inferred from the appropriate $Y^2$ isochrone, suggests we should adopt a slightly more conservative upper limit on {\it v} sin {\it i}.  We have therefore converted all values from Paper 1 up to 20 km s$^{-1}$ into upper limits at 20 km s$^{-1}$.

\begin{longrotatetable}
	\begin{deluxetable*}{cccccccccccccc}
		\label{tab:new_parameter}
		\tablecaption{Stellar parameters of new M48 candidates}
		\tabletypesize{\tiny}
		\setlength{\tabcolsep}{3pt}
		\decimalcolnumbers
		\renewcommand{\arraystretch}{1.0}
		\tablehead{
			\colhead{Old Id} & \colhead{Star Id} & \colhead{RA} & \colhead{DEC} &
			\colhead{$V^1$} & \colhead{$B-V^1$} &
			\colhead{$V_{\rm{RAD}}^2$} & \colhead{$\sigma^2$} &
			\colhead{$V_{\rm{RAD}}^3$} & \colhead{$\sigma^3$} &
			\colhead{{\it v} sin{\it i}$^4$} & \colhead{$\sigma^4$} &
		    \colhead{mm$^{5}$} &  \colhead{configuration$^{6}$} \\
			 \colhead{} &  \colhead{} &h m s&$\arcdeg\ \arcmin\ \arcsec$& mag & mag & \colhead{km s$^{-1}$} & \colhead{km s$^{-1}$} & \colhead{km s$^{-1}$} & \colhead{km s$^{-1}$} &
			\colhead{km s$^{-1}$} & \colhead{km s$^{-1}$} & \colhead{} & \colhead{}
		} 
		\startdata
		\hline
		--& 2119 &	8 11 59.96 & -5 57 33.8	& 12.86	 &  0.400 &  7.53,10.43	        &  6.96,2.51	    &   11.96 &	 4.45	&  49	& 3.5  & sm	& m48gf2, m48gf3 \\
		2014 & 3001 &	8 13  5.38 & -5 45  0.5	& 9.478	 &  0.031 &	28.44,30.08	        &  2.11,4.44	    &   -- &	-- 	&  52	& 14.5  & bm	& m48gb          \\
		2019 & 3002 &	8 13  4.96 & -5 53  4.8	& 9.935	 &  0.094 &	22.01,-13.19        &  5.03,9.65	    &  -- &	 -- 	&  250	& --  & sm	& m48gb          \\
		2038 & 3003 &	8 13  9.50 & -5 27  1.1	& 10.625 &	9.999  &	24.14,21.70	        &  2.36,2.46	    &  -14.53 &	18.12	& 69	& 13.5 & um & m48gb          \\
		2049 & 3004 &	8 14  2.61 & -5 24 17.0	& 10.977 &	9.999  &	17.79,17.57	        &  6.20,3.98	    &  -- &	-- 	&  42	& 10.7  & um	& m48gb          \\
		-- & 3005 &	8 14 21.78 & -5 47 23.2	& 10.988 &	0.240 &	22.29,14.16	        &  7.90,2.96	    &   -- &	-- 	&  56	& 10.3  & bm	& m48gb          \\
		2060 & 3006 &	8 14 15.14 & -5 16 55.2	& 11.302 &	0.226  &	37.02,32.97	        &  2.91,2.44	    &   -- &	 -- 	&  23	& 3.2  & bm$^*$& m48gb          \\
		2061 & 3007 &	8 14 56.37 & -5 40 32.7	& 11.332 &	0.185  &	13.78,8.92	        &  8.57,6.21	    &   -- &	-- 	&  58	& 13.5  & um & m48gb          \\
		--& 3008 &	8 11 56.01 & -5 29 25.0	& 11.367 &	0.152 &	-6.07,0.00	        &  1.65,1.62	    &   -- &	 -- 	&  67	& 5.7  & bm	& m48gb          \\
		-- & 3010 &	8 12 38.95 & -5 41 51.8	& 11.633 &	0.302 &	9.06,6.75	        &  4.28,3.23	    &    7.75 &	 4.17	&  53	& 3.3  & sm	& m48gb          \\
		2076 & 3011 &	8 14 41.19 & -5 23 45.3	& 11.671 &	0.235  &	9.71,8.07	        &  0.87,1.32	    &    2.91 &	 7.44	&  67	& 4.6  & sm	& m48gb          \\
		2080 & 3012 &	8 13 47.70 & -5 46  0.7	& 11.701 &	0.245 &	7.61,8.80	        &  2.17,1.46 &     7.5 &	 7.15	&  61	& 5.3  & sm	& m48gb \\
		2091 & 3013 &	8 13  2.71 & -5 58 59.5	& 11.944 &  0.291 &	9.20,8.95	        &  1.90,2.21	    &     9.3 &	 2.45	&  32	& 1.9  & sm	& m48gb          \\
		2095 &3015 &	8 14 49.66 & -5 18 41.7	& 12.082 &	0.300  &	11.66,8.27	        &  7.55,6.69	    &    8.43 &	 6.57	&  61	& 5.0  & sm$^*$	& m48gf1         \\
		--&3016 &	8 13 31.93 & -5 49 29.6	& 12.358 &	0.463 &	7.98,8.17	        &  3.37,1.35	    &    -- &	 -- 	&  30	& 1.0  & bm	& m48gf2, m48gf3 \\
		--&3018 &	8 14  1.95 & -5 46 50.9	& 12.406 &	0.497 & 29.23,-37.57	    &  5.07,9.21	    &  -- &	-- 	&  49	& 8.5  & bm	& m48gf1         \\
		--&3019 &	8 13 28.98 & -6  0 50.9	& 12.51  &	0.419 &	9.54,9.75	        &  1.03,0.97	    &    9.85 &	 1.06	&  <20	& 0.7  & um & m48gf1         \\
		2115&3020 &	8 12 53.59 & -5 32 23.5	& 12.783 &	0.435 & 11.51,9.81,9.54	    &  2.69,5.81,3.94	&   10.06 &	 3.80	&  41	& 2.9  & sm	& m48gf1, m48gf3 \\
		--&3021 &	8 12  8.73 & -6  4 16.5	& 13.058 &	0.446 &	9.83,10.03	        &  7.55,0.88	    &    8.93 &	 6.92	&  49	& 5.5  & sm	& m48gf2, m48gf3 \\
		--&3022 &	8 13 45.61 & -5 35  3.3	& 13.139 &	0.643 &	34.09,21.46	        &  3.68,1.83	    &  -- &	 -- 	&  28	& 1.4  & bm	& m48gf1         \\
		2125&3023 &	8 14 25.25 & -5 33 50.3	& 13.192 &	0.473 &	9.89,9.44	        &  1.40,1.72	    &    9.56 &	 1.90	&  <20	& 1.2  & sm	& m48gf1         \\
		--&3024 &	8 14 18.81 & -5 41 32.8	& 13.215 &	0.589 &	6.44,6.00	        &  2.70,2.32	    &    6.21 &	 2.46	&  34	& 1.9  & um & m48gf1         \\
		2126&3025 &	8 12 51.02 & -5 58 55.6	& 13.199 &	0.475 &	9.12,7.97	        &  1.99,1.16	    &    8.37 &	 1.17	&  21	& 0.7  & sm	& m48gf2,m48gf3  \\
		--&3026 &	8 12 42.22 & -5 43 27.8	& 13.223 &	0.599 &	-45.58,12.28	    &  2.71,5.07	    &   -- &	 -- 	&  71	& 4.6  & bm	& m48gf1         \\
		2130&3027 &	8 14 56.28 & -5 38 26.6	& 13.279 &	0.496  &	7.24,8.40	        &  2.36,2.45	    &    8.46 &	 2.73	&  32	& 2.0  & sm	& m48gf1         \\
		2135&3028 &	8 14 59.35 & -5 21 49.8	& 13.411 &  0.51  &	8.37,9.15	        &  1.76,1.29	    &    9.23 &	 1.50	&  20	& 0.9  & sm	& m48gf1         \\
		--&3029 &	8 12 33.52 & -5 58 12.9	& 13.415 &	0.474 &	   7.69	            &     1.13	        &     7.6 &	 1.32	&  <20	& 0.6  & sm	& m48gf2         \\
		--&3030 &	8 13 51.71 & -5 53  4.0	& 13.522 &	0.557 &	8.15, 9.54	        &  2.75, 2.13	    &    8.56 &	 2.07	&  29	& 1.6  & um & m48gf2, m48gf3 \\
		--&3031 &	8 12  8.87 & -5 38  6.8	& 13.544 &	0.648 &	8.11,8.10	        &  1.03,0.57	    &    8.23 &	 0.58	&  21	& 0.4  & um & m48gf1         \\
		2140&3032 &	8 13 57.34 & -5 26 49.0	& 13.605 &	0.541  &	8.30,8.99	        &  1.55,1.18	    &    9.07 &	 1.29	&  21	& 0.8  & sm	& m48gf1         \\
		--&3033 &	8 13 58.36 & -5 42 33.4	& 13.609 &	0.643 &	8.63, 38.02	        &  1.19, 1.45	    &  -- &	 -- &  45	& 1.6  & bm	& m48gf2, m48gf3 \\
		2141&3034 &	8 12 45.29 & -5 55 47.0	& 13.609 &	0.543 & -7.07,-14.02,6.65   &  2.64,1.76,2.96	&  -12.07 &	 1.79	&  30	& 1.4  & ulm & m48gb, m48gf3  \\
		--&3035 &	8 13 26.17 & -5 34 28.6	& 13.693 &	0.518 &	28.38,0.26	        &  1.20,1.17        &   1.13 &	 1.43	&  <20	& 0.9  & ulm & m48gf1         \\
		--&3036 &	8 13 20.58 & -6  1  3.3	& 13.698 &	0.56  &	7.31,7.01	        &  1.04,1.01        &   7.17 &	 1.05	&  <20	& 0.6  & sm	& m48gf1         \\
		--&3037 &	8 13 29.04 & -5 17 52.2	& 13.744 &	0.565 &	7.44,7.85	        &  3.18,1.24        &   7.93 &	 1.43	&  <20	& 0.8  & sm	& m48gf1         \\
		2148&3038 &	8 14 56.43 & -5 32 11.9	& 13.748 &	0.584  &	8.40,8.29	        &  2.06,1.28        &   8.41 &	 1.39	&  <20	& 0.9  & sm	& m48gf1         \\
		--&3039 &	8 12 34.83 & -5 35 59.5	& 13.760 & 0.675  &	8.05,8.61	        &  0.57,0.71        &   -- &	 -- 	&  24	& 1.7  & bm	& m48gf1         \\
		--&3040 &	8 11 58.20 & -5 32  7.6	& 13.838 &	0.58  &	8.43,8.08	        &  1.90,1.29        &   8.22 &	 1.30	&  <20	& 0.8  & sm	& m48gf1         \\
		--&3041 &	8 14 49.46 & -5 30 46.4	& 13.869 &	9.999  &	12.61,43.29	        &  1.66,1.41	    &  -- &	 -- &  32	& 1.6  & bm	& m48gf2, m48gf3 \\
		--&3042 &	8 13 31.10 & -5 26 16.5	& 13.894 &	0.59  &	8.5,9.23	        &  1.68,1.11	    &    9.3 &	 1.23	&  <20	& 0.6  & um & m48gf1         \\
		--&3043 &	8 15 27.34 & -5 58 38.6	& 14.068 &	0.628 &	    7.83	        &    0.69	        &   7.81 &	 0.65	&  <20	& 0.2  & sm	& m48gf2         \\
		--&3044 &	8 14  8.79 & -5 43 41.5	& 14.089 &	0.755 &	101.19,83.86	    &  2.31,3.56        &  -- &	 -- &  37	& 2.5  & bm	& m48gf2, m48gf3 \\
		--&3045 &	8 13 23.12 & -5 41 17.2	& 14.109 &	0.735 &	-9.73,-11.23	    &  8.80,2.89        & -- &	-- 	&  52	& 5.1  & bm	& m48gf1         \\
		--&3046 &	8 13 39.42 & -5 49 35.9	& 14.174 &	0.894 &	3.59,6.30	        &  0.97,1.82	    &  -- &	 -- &  50	& 1.9  & bm	& m48gf1         \\
		--&3047 &	8 13 24.44 & -6  1 52.8	& 14.192 &	0.621 &	7.71,8.60	        &  0.84,0.51	    &   8.63 &	 0.56	&  <20	& 0.3  & sm	& m48gf1         \\
		2165&3048 &	8 13 46.52 & -5 38 53.3	& 14.225 &	0.636 &	9.35,8.57	        &  1.10,0.75	    &   8.79 &	 0.58	&  <20	& 0.3  & sm	& m48gf1         \\
		--&3049 &	8 13 45.58 & -5 48 42.0	& 14.399 &	0.665 &	6.44,8.03,7.62	    &  2.70,2.42,1.19	&   7.98 &	 0.91	&  <20	& 0.6  & sm	& m48gf1, m48gf3 \\
		--&3050 &	8 12 52.18 & -6  1 31.0	& 14.447 &	0.66  &	8.12,9.29           &  0.95,0.61	    &   9.29 &	 0.55	&  <20	& 0.2  & sm	& m48gf1         \\
		2178&3051 &	8 14 55.65 & -5 35 38.6	& 14.542 &	0.694  &	   8.65	            &     0.55	        &   8.57 &	 0.56	&  <20	& 0.3  & sm	& m48gf2         \\
		--&3052 &	8 14  6.92 & -5 45 26.0	& 14.574 &	0.809 &	7.19,6.84	        &  1.43,0.54	    &   -- &	 -- 	&  <20	& 0.3  & bm	& m48gf1         \\
		--&3053 &	8 12 33.01 & -5 46 46.8	& 14.682 &	0.699 &	7.83,8.19	        &  0.83,0.45	    &   8.28 &	 0.41	&  <20	& 0.2  & sm	& m48gf1         \\
		--&3054 &	8 13 16.07 & -5 16 46.7	& 14.786 &	0.724 &	5.13,8.15	        &  2.00,0.64	    &   8.05 &	 0.65	&  <20	& 0.2  & sm	& m48gf1         \\
		--&3055 &	8 13 36.75 & -6  0  4.7	& 14.842 &	0.797 &	7.31,8.18	        &  0.98,0.51	    &   8.21 &	 0.42	&  <20	& 0.2  & sm	& m48gf1         \\
		--&3056 &	8 14  5.37 & -5 47 13.0	& 14.844 &	0.825 &	8.71,8.95	        &  0.93,0.41	    &   9.05 &	 0.53	&  <20	& 0.3  & um & m48gf1         \\
		2203& 3057 &	8 14 54.97 & -5 35 50.4	& 14.894 &	0.770  &	6.86,8.68	        &  1.33,0.44	    &   8.70 &	 0.42	&  <20	& 0.2  & sm	& m48gf1         \\
		--&3058 &	8 13  9.27 & -6 11 15.3	& 15.055 &	0.789 &	6.46, 7.77	        &  0.88,0.54	    &   7.32 &	 0.56	& <20	& 0.3  & sm	& m48gf2, m48gf3 \\
		--&3059 &	8 12 51.32 & -5 27 44.4	& 15.118 &	0.802 &	7.14,7.05	        &  1.45,0.47	    &   7.18 &	 0.50	&  <20	& 0.2  & sm	& m48gf1         \\
		--&3060 &	8 12 37.27 & -5 47 52.3	& 15.221 &	0.777 &	9.26,9.02	        &  0.99,0.45	    &   9.17 &	 0.43	&  <20	& 0.2  & sm	& m48gf1         \\
		--&3061 &	8 13 25.25 & -5 54 36.9	& 15.32	 &  0.877 & -11.37,-10.38,20.89 & 	1.10,0.65,0.84	&   9.07 &	 1.71	&  43	& 1.3  & sln & m48gf1, m48gf3 \\
		2232& 3062 &	8 14  9.58 & -5 43 40.8	& 15.33	 &  0.852  &	8.9, 9.33 &  1.01,0.41	    &   9.42 &	 0.46	&  <20	& 0.2  & sm	& m48gf1         \\
		--&3063 &	8 12 50.38 & -5 37 49.3	& 15.33	 &  0.86  &	8.15,7.54	        &  1.18,0.43	    &   7.71 &	 0.44	&  <20	& 0.2  & sm	& m48gf1         \\
		2235 & 3064 &	8 14  3.31 & -5 34  7.9	& 15.386 &	0.857 &	10.01,8.60	        &  1.22,0.34	    &   8.78 &	 0.32	&  <20	& 0.2  & sm	& m48gf1         \\
		2238&3065 &	8 14  2.84 & -5 21  6.2	& 15.53	 &  0.900  &	7.51,8.25	        &  1.09,0.47	    &   8.33 &	 0.47	&  <20	& 0.2  & sm	& m48gf1         \\
		--&3066 &	8 12 45.59 & -5 16 50.7	& 15.562 &	0.89  &	8.21,6.79	        &  1.15,0.77	    &   7.22 &	 0.90	&  <20	& 0.2  & um & m48gf1         \\
		--&3067 &	8 13 51.54 & -5 24 32.0	& 15.856 &	9.999  &	12.32,8.36	        &  3.46,0.79	    &   8.55 &	 0.76	&  <20	& 0.4  & um & m48gf1         \\
		2267&3068 &	8 14 57.19 & -5 26  7.9	& 15.954 &	1.013  &	7.41	            &   0.92	        &   7.92 &	 0.95	&  <20	& 0.4  & sm	& m48gf3         \\
		--&3069 &	8 13 43.30 & -5 18 31.1	& 15.94	 &  0.973 &	8.88,8.29,7.85	    & 1.31,0.64,0.79	&   8.34 &	 0.62	&  <20	& 0.3  & sm	& m48gf1, m48gf3 \\
		--&3070 &	8 14 32.97 & -5 35 20.9	& 15.966 &	0.992 &	7.4,7.64,9.03	    & 3.17,0.56,0.65	&   8.38 &	 0.47	&  <20	& 0.3  & um & m48gf1, m48gf3 \\
		2274&3071 &	8 14 30.11 & -5 30 38.0	& 16.022 &	0.985 &	7.42,8.4	        & 0.71,0.48	        &   8.13 &	 0.50	&  <20	& 0.3  & sm	& m48gf2, m48gf3 \\
		2281&3073 &	8 14 22.19 & -5 43  5.3	& 16.09	 & 1.014  &	7.97,8.54,7.45	    & 1.44,0.50,0.70	&   8.44 &	 0.50	&  <20	& 0.3  & sm	& m48gf1, m48gf3 \\
		\hline
		\enddata
		\tablecomments{1. $V$ magnitude and $B-V$ color from our M48 photometry.
			2. Radial velocity ($V_{\rm{RAD}}$) and error in km s$^{-1}$, reported for individual nights.
			3. $V_{\rm{RAD}}$ and error in km s$^{-1}$ measured by using the combined spectra for single stars and stars with uncertain multiplicity. We do not report the combined $V_{\rm{RAD}}$  for binaries.
			4. Rotational velocity ({\it v} sin{\it i}) and errors in km s$^{-1}$. 
			5. Multiplicity \& membership (mm) determination. The same designations are used as in Paper 1 (\citealt{2020AJ....159..220S}). * means either multiplicity or membership has been changed from Paper I.
			6. Configuration(s) that include this star.}
	\end{deluxetable*}
\end{longrotatetable}

In total, among the 42 new stars, 18 are single members, 13 are binary members, 9 are members of uncertain multiplicity, 1 is a likely member of uncertain multiplicity, and 1 is a single likely non-member. In view of the new data, we change star 2060 from ulm to bm, and star 2095 from sln to sm. Combined with the 287 stars of Paper 1, Table \ref{tab:tally} summarizes multiplicity/membership for our total sample of 329 stars.

\begin{deluxetable}{ccc}
	\label{tab:tally}
	\tablecaption{Multiplicity/membership tally}
	\tabletypesize{\scriptsize}
	\tablehead{
		\colhead{category} &
		\colhead{number of stars} &
		\colhead{description} \\
	} 
	\startdata
	\hline
	sm & 171 & single members \\ 
	slm & 2 & single likely members \\
	sln & 28 & single likely non-members \\
	sn  & 56 & single non-members \\
	bm & 25 & binary members \\
	blm & 1 & binary likely members \\
	bln & 0 & binary likely non-members \\
	bn & 5 & binary non-members \\
	um & 25 & members of uncertain multiplicity \\
	ulm & 10 & likely members of uncertain multiplicity \\
	uln & 3 & likely non-members of uncertain multiplicity \\
	un & 3 & non-members of uncertain multiplicity \\
	\hline
	\enddata
	\vspace{-10mm}
\end{deluxetable}

Hereafter, we restrict attention to the 234 members and likely members (``m" and ``lm") of M48, all of which appear in Table \ref{tab:candidates}.

\begin{longrotatetable}
	\begin{deluxetable*}{ccccccccccccccccccccc}
		\label{tab:candidates}
		\tablecaption{Stellar atmosphere and Li abundances for M48 members and likely members}
		\tabletypesize{\tiny}
		\setlength{\tabcolsep}{3pt}
		\decimalcolnumbers
		\renewcommand{\arraystretch}{1.0}
		\tablehead{
			\colhead{Star Id} &\colhead{WOCS Id$^1$} & \colhead{mm} &
			\colhead{$(B-V)_{\rm{eff}}^2$} & \colhead{$\sigma^2$} & \colhead{{\it v} sin {\it i}}$^3$ &  \colhead{$\sigma^3$} &  \colhead{$H_{\alpha}^3$} &  \colhead{Period$^4$} & \colhead{$e_P^4$} &  \colhead{TESS Period$^5$} & \colhead{$e_P^5$} & $T_{\rm eff}^6$ & \colhead{$\sigma^6$} & \colhead{log {\it g}$^6$} & \colhead{$V_{\rm t}^6$} & \colhead{SNR$^7$} & \colhead{FWHM$^7$} & \colhead{A(Li)} & \colhead{$\sigma^8$} & \colhead{comments} \\
			\colhead{} & \colhead{} &\colhead{} &
			\colhead{mag} & \colhead{mag} & \colhead{km s$^{-1}$} &  \colhead{km s$^{-1}$} &  \colhead{} & \colhead{day} & \colhead{day} & \colhead{day} & \colhead{day} & K & K &  & \colhead{km s$^{-1}$} & \colhead{} & \colhead{m$\AA$} & \colhead{dex} & \colhead{dex} & \colhead{} \\
		} 
		\startdata
		\hline
		 2002 &	1010 & slm &0.768 &0.002&<20& 0.4& no	& --   & --   &  --  &  -- & 5520 & 6	 &   2.83 &	1.13 &	1063 &  858	  & 1.9 & 0.02 & giant \\
		2003 &	1025&sm &	1.067 &	0.002 &	<20& 0.6	  & no	& --   & --   &  --  &  -- & 4752 & 5	 &   2.44 &	1.18 &	959	 &  894	  &  1.0 & 0.02 & giant \\
		2004 &	1020&sm &	0.934 &	0.004 &	21	   & 0.6  & no	& --   & --   &  --  &  -- & 5071 & 11	 &   2.77 &	1.14 &	904	 &  999	  & 1.35 & 0.02 & giant \\
		2010 &	1011&sm &	0.057 &	0.010 &	34	   & 4.2  & no	& --   & --   &  --  &  -- & 8537 & 51	 &   3.78 &	4.21 &	518	 &  1273  &	 <3.6  & -- & \\
		2011 &	2019&sm &	0.063 &	0.012 &	300	   & --	  & yes	& --   & --   &  --  &  -- & 8505 & 61	 &   3.83 &	4.12 &	807	 &  13416 & <3.45  & -- & \\
		2012 &	1005&bm &	0.073 &	0.008$^*$&  --& --	  & no	& --   & --   & 1.73  & 0.09 & 8453 & --	 &   3.90 &	3.99 &	740	 &  3160  & <3.5 & -- & \\
		2014 &	1018& bm &	0.076 &	0.009 &	52	& 14.5    & no	& --   & --   &  --  &  -- & 8437 & 45	 &   3.92 &	3.95 &	636	 &  3026  & <3.5 & -- & \\
		2015 &2011&	bm &0.072 &	0.009 & 29(31) & 3.2(5.1) & no	& --   & --   & 4.34  & 0.32 & 8460 & 45	 &   3.89 &	4.01 &	661	 &  887	  & <3.3 & -- & \\
		2016 &	1028& blm &0.084 &0.009 &<20&4.8 & no	& --   & --   & 3.06  & 0.83 & 8395 & 45	 &   3.97 &	3.85 &	816	 &  1110  & <2.8 & -- & \\
		2017 &	1009&sm &	0.066 &	0.007 &	23	   & 2.5  & no	& --   & --   &  --  &  -- & 8490 & 35	 &   3.85 &	4.08 &	579	 &  1009  &	 <3.2 & -- & \\
		2018 &	2008&bm &	0.141 &	0.029 &	--	   & --	  & no	& --   & --   &  --  &  -- & 8104 & 140 &   4.18 &	3.35 &	664	 &  4163  &	 <2.9 & -- & \\
		2019 &	2026&sm &	0.089 &	0.006 &	250	   & --	  & yes	& --   & --   &  --  &  -- & 8368 & 30	 &   4.00 &	3.79 &	561	 &  11180 & <3.45  & -- & \\
		2022 &	2005&sm &	0.047 &	0.006 &	230	   & --	  & yes	& --   & --   &  --  &  -- & 8575 & 31	 &   3.69 &	4.36 &	515	 &  10285 & <3.65 & -- & \\
		2023 &	1003&sm &	0.109 &	0.013 &	150	   & --	  & yes	& --   & --   &  --  &  -- & 8263 & 64	 &   4.09 &	3.60 &	570	 &  6708  & <3.5 & -- & \\
		2025 &	1013&bm &	0.042 &	0.008 &	40	   & 6	  & no	& --   & --   &  --  &  -- & 8575 & 41	 &   3.64 &	4.43 &	513	 &  1689  & <3.3 & -- & \\
		2029 &	-- & sm &	0.063 &	0.008 &	230	   & --	  & yes	& --   & --   &  --  &  -- & 8504 & 40	 &   3.83 &	4.12 &	376	 &  10285 & <3.75  & -- & \\
		2030 &	1022&sm &	0.068 &	0.008 &	42	   & 6.5  & no	& --   & --   &  --  &  -- & 8480 & 40	 &   3.87 &	4.06 &	400	 &  1568  & <3.4 & -- & \\
		2032 &	2729&sm &	0.074 &	0.006 &	<20	   & 1.9  & no	& --   & --   &  --  &  -- & 8449 & 30	 &   3.91 &	3.98 &	376	 &  1156  & <3.1  & -- & \\
		2033 &	3006&bm &	0.082 &	0.005$^*$  & 52& 8.3  & no	& --   & --   & 4.54  & 0.31 & 8406 & --	 &   3.96 &	3.88 &	433	 &  2138  &	 <3.3 & -- & \\
		2035 &	2023&sm &	0.149 &	0.014 &	280	   & --	  & yes	& --   & --   &  --  &  -- & 8062 & 67	 &   4.19 &	3.30 &	404	 &  12521 &	<3.4 & -- & \\
		2036 &	3729&sm &	0.092 &	0.012 &	230	   & --	  & yes	& --   & --   &  --  &  -- & 8351 & 60	 &   4.02 &	3.76 &	447	 &  10285 &	<3.5  & -- & \\
		2037 &	3023 & bm &	0.085 &	0.015 &	24	   & 2.1  & no	& --   & --   &  --  &  -- & 8390 & 75	 &   3.98 &	3.84 &	437	 &  1217  &	 <3.3  & -- & \\
		2038 &	2737 & um &0.136 &0.013 &69& 13.5 & no	& --   & --   &  --  &  -- & 8127 & 63	 &   4.17 &	3.38 &	500	 &  2557  &	 <3.2 & -- & \\
		2039 &	1016&sm &	0.070 &	0.018 &	60	   & --	  & no	& --   & --   &  --  &  -- & 8470 & 90	 &   3.88 &	4.03 &	379	 &  2859  &	 <3.7  & -- & \\
		2040 &	2014 & um &0.094 &	0.010 &	--&-- & no	& --   & --   &  --  &  -- & 8343 & 50	 &   4.02 &	3.74 &	764	 &  3560  &	<3.2 & -- & \\
		2041 &	2018& sm &	0.195 &	0.017 &	67	   & 8.4  & no	& --   & --   &  --  &  -- & 7832 & 80	 &   4.24 &	3.05 &	363	 &  2092  &	 <3.3 & -- & \\
		2042 &	1746&sm &	0.165 &	0.015 &	280	   & --	  & yes	& --   & --   &  --  &  -- & 7980 & 71	 &   4.22 &	3.20 &	386	 &  12521 &	<3.3  & -- & \\
		2043 &	1730&sm &	0.197 &	0.017 &	130	   & --	  & yes	& --   & --   &  --  &  -- & 7824 & 80	 &   4.24 &	3.05 &	376	 &  5813  &	 <3.4  & -- & \\
		2044 &	2730& um &0.106 &0.016&79& 13.2  & no	& --   & --   &  --  &  -- & 8281 & 79	 &   4.08 &	3.63 &	573	 &  3755  &	 <3.3 & -- & \\
		2045 &	1735&sm &	0.108 &	0.007 &	70	   & --	  & yes	& --   & --   &  --  &  -- & 8270 & 34	 &   4.08 &	3.61 &	401	 &  3130  &	 <3.4 & -- & \\
		2046 &	2742&sm &	0.094 &	0.011 &	200	   & --	  & yes	& --   & --   &  --  &  -- & 8345 & 55	 &   4.02 &	3.75 &	367	 &  8944  &	 <4.3 & -- & \\
		2048 &	3015&sm &	0.107 &	0.013 &	50	   & 7.7  & no	& --   & --   & 3.46  & 0.29 & 8274 & 64	 &   4.08 &	3.61 &	316	 &  1669  &	 <3.4 & -- & \\
		2049 &	2741& um &	0.170 &	0.010 &	250&--& yes	& --   & --   &  --  &  -- & 7956 & 48	 &   4.22 &	3.18 &	485	 &  1019  &	 <3.2  & -- & \\
		2050 &	3007& ulm &0.177 &	0.008 &	--& --& no	& --   & --   & 7.31  & 0.69 & 7921 & 38	 &   4.23 &	3.14 &	98	 &  1758  &	 <3.45 & -- & \\
		2053 &	3026& um &	0.195 &	0.006 &	57& 7.0 & no	& --   & --   & 3.19  & 0.13 & 7832 & 28	 &   4.24 &	3.05 &	587	 &  2035  &	 <3.3 & -- & \\
		2054 &	3736&sm &	0.131 &	0.010 &	180	   & --	  & yes	& --   & --   &  --  &  -- & 8154 & 49	 &   4.15 &	3.42 &	373	 &  8049  &	 <3.8 & -- & \\
		2057 &	2016&sm &	0.225 &	0.010 &	150	   & --	  & yes	& --   & --   &  --  &  -- & 7683 & 46	 &   4.25 &	2.92 &	353	 &  6708  &	<3.3 & --& \\
		2058 &	4026&sm &	0.188 &	0.006 &	220	   & --	  & yes	& --   & --   &  --  &  -- & 7868 & 28	 &   4.24 &	3.09 &	318	 &  9838  & <3.4 & --& \\
		2060 &	3757& bm$^*$ &	0.208 &	0.026 &	<20& 3.8  & no	& --   & --   &  --  &  -- & 7768 & 121 &	 4.25 &	3.00 &	627	 &  604	  & <3.3  & --& \\
		2061 &	1740& um &	0.210 &	0.004 &	--&-- & no	& --   & --   &  --  &  -- & 7758 & 19	 &   4.25 &	2.99 &	477	 &  2384  &	 <3.3 & --  & \\
		2063 &	2024& um &	0.186 &0.022 &81&11.4 & no	& --   & --   & 7.42  & 0.32 & 7876 & 104 &	 4.23 &	3.10 &	521	 &  2023  &	<3.2 & -- & \\
		2064 &	4015&sm &	0.188 &	0.017 &	90	   & --	  & yes	& --   & --   &  --  &  -- & 7865 & 80	 &   4.24 &	3.09 &	410	 &  4025  & <3.3 & -- & \\
		2068 &	3018&sm &	0.207 &	0.013 &	77	   & 0.4  & no	& --   & --   &  --  &  -- & 7774 & 60	 &   4.24 &	3.00 &	388	 &  2885  &	<3.3 & -- & \\
		2070 &	3008& ulm &0.245&0.003$^*$&32&5.4 & no	& --   & --   & 1.73  & 0.44 & 7589 & --	 &   4.25 &	2.84 &	305	 &  1173  &	<3.3  & -- & \\
		2073 &	3732& ulm &0.248&0.012 &--& --	  & no	& --   & --   &  --  &  -- & 7574 & 54	 &   4.25 &	2.83 &	139	 &  4036  & -- & -- & spectrum not good \\
		2074 &	2753& ulm &0.222&0.018 &55& 8.0	  & no	& --   & --   &  --  &  -- & 7700 & 83	 &   4.25 &	2.94 &	347	 &  2177  &	<3.2 & -- & \\
		2076 &	2751&sm &	0.204 &	0.012 &	67	   & 4.6  & no	& --   & --   &  --  &  -- & 7786 & 56	 &   4.24 &	3.01 &	461	 &  1552  & <3.3  & -- & \\
		2079 &	2013& um &	0.228 &	0.017 &	--&-- & no	& --   & --   &  --  &  -- & 7670 & 78	 &   4.25 &	2.91 &	204	 &  1685  & <3.6  & -- & \\
		2080 &	4007&sm &	0.228 &	0.014 &	60	   & 3.34 & no	& --   & --   &  --  &  -- & 7669 & 64	 &   4.25 &	2.91 &	639	 &  2248  & 3.65 & 0.03 & \\
		2082 &	3011& um &	0.221 &	0.011 &	--&-- & no	& --   & --   &  --  &  -- & 7704 & 51	 &   4.25 &	2.94 &	386	 &  688	  &  <3.0  & --& \\
		2083 &	2022&sm &	0.248 &	0.011 &	210	   & --	  & yes	& --   & --   &  --  &  -- & 7576 & 50	 &   4.25 &	2.84 &	327	 &  9391  & <3.6  & --& \\
		2085 &	4733& ulm &0.258 &0.020&47& 0.3	  & no	& --   & --   &  --  &  -- & 7527 & 90	 &   4.25 &	2.80 &	128	 &  1345  & <3.4  & --& \\
		2086 &	1738&bm &	0.277 &	0.022 &	--	   & --	  & no	& --   & --   &  --  &  -- & 7439 & 98	 &   4.25 &	2.73 &	375	 &  3178  & <3.3 & --& \\
		2088 &	4737& um &0.306&0.036&--& --	  & no	& --   & --   &  --  &  -- & 7302 & 157 &	4.25 &	2.62 &	135	 &  6729  & <3.8 & -- & \\
		2089 &	4016& um &0.265&0.013&--& --	  & no	& --   & --   &  --  &  -- & 7493 & 58	 &   4.25 &	2.77 &	463	 &  945	  &  <3.1 & --& \\
		2091 &	4736& sm &	0.304 &	0.024 &	33	& 1.35	  & no	& --   & --   &  --  &  -- & 7313 & 105 &	4.25 &	2.63 &	596	 &  1221  &	3.73 & 0.03 & \\
		2092 &	2745& um &	0.275&0.025&--& --	  & no	& --   & --   &  --  &  -- & 7446 & 112 &	4.25 &	2.73 &	530	 &  679	  &  <2.7  & -- & \\
		2095 &	-- & sm$^*$	 &	0.274 &	0.022 &	61& 5.0	  & no	& --   & --   &  --  &  -- & 7443 & 99	 &   4.25 &	2.73 &	534	 &  2252  & 3.1 & 0.05 & \\
		2096 &	5008&sm &	0.335 &	0.014 &	38	& 2.4	  & no	& --   & --   &  --  &  -- & 7170 & 60	 &   4.25 &	2.51 &	332	 &  1552  &	2.85 & 0.05 & \\
		2098 &	5015& um &	0.317 &	0.015 &	--&-- & no	& --   & --   &  --  &  -- & 7251 & 65	 &   4.25 &	2.57 &	493	 &  2651  &	 <3.3  & -- & \\
		2099 &	4018& ulm &0.355 &	0.010 &27&3.6 & no	& --   & --   & 2.52  & 0.68 & 7079 & 42	 &   4.26 &	2.43 &	335	 &  1566  &	 <3.4  & -- & \\
		2100 &	4752& ulm &0.360 &	0.024 &21&1.4 & no	& --   & --   & 2.27  & 0.23 & 7056 & 101 &	4.26 &	2.41 &	303	 &  945	  & 3.12  & 0.07 & \\
		2101 &	2021& bm &	0.336 &	0.017 &	92     & 11.7 & no	& --   & --   &  --  &  -- & 7165 & 73	 &   4.25 &	2.50 &	258	 &  3838  &	 <3.3  & -- & \\
		2102 &	3741& um &	0.345 &0.024 &66& 5.1 & no	& --   & --   &  --  &  -- & 7124 & 102 &	4.25 &	2.47 &	293	 &  2512  &	 <3.1  & --& \\
		2103 &	5016& ulm &0.349 &	0.017 &50&7.4 & no	& --   & --   &  --  &  -- & 7106 & 72	 &   4.25 &	2.45 &	96	 &  1649  &	<3.2  & --& \\
		2104 &	4732& sm &	0.335 &	0.028 &	41	   & 3.8  & no	& --   & --   &  --  &  -- & 7170 & 120 &	4.25 &	2.51 &	349	 &  1441  &	 3.25  & 0.06 & \\
		2105 &	7025& um &	0.355 &0.012&57& 5.8  & no	& --   & --   &  --  &  -- & 7079 & 51	 &   4.26 &	2.43 &	461	 &  3033  &	 3.3 & 0.05 & \\
		2108 &	6754&sm &	0.395 &	0.016 &	27	   & 1.2  & no	& --   & --   &  --  &  -- & 6903 & 66	 &   4.27 &	2.27 &	227	 &  1056  &	 3.3 & 0.06 & \\
		2113 &	5739&sm &	0.400 &	0.015 &	31	   & 1.3  & no	& --   & --   &  --  &  -- & 6881 & 62	 &   4.27 &	2.25 &	258	 &  1506  &	 2.9 & 0.06 & \\
		2115 &	6732&sm &	0.436 &	0.018 &	40	   & 1.8  & no	& --   & --   & 5.73  & 0.09 & 6724 & 72	 &   4.29 &	2.10 &	265	 &  1386  &	 <2.6  & --& \\
		2119 &	9757&sm &	0.410 &	0.013 &	49     & 3.54 & no	& --   & --   &  --  &  -- & 6836 & 53	 &   4.28 &	2.21 &	239	 &  1750  &	<2.5  & --& \\
		2120 &	7752&sm &	0.456 &	0.021 &	31	   & 3.8  & no	& --   & --   &  --  &  -- & 6642 & 83	 &   4.30 &	2.02 &	261	 &  1135  &	 2.6 & 0.08 & \\
		2121 &	7741&sm &	0.452 &	0.013 &	41	   & 1.6  & no	& --   & --   & 2.88  & 0.21 & 6658 & 52	 &   4.30 &	2.04 &	237	 &  1388  &	2.2 & 0.17 & \\
		2122 &	8745& um &0.451&0.011&86& 4.2	  & no	& --   & --   & 3.01  & 0.26 & 6661 & 44	 &   4.30 &	2.04 &	315	 &  1874  & <2.5 & -- & \\
		2125 &	6731&sm &	0.473 &	0.010 &	<20	   & 0.7  & no	& --   & --   &  2.8  & 0.15 & 6568 & 39	 &   4.31 &	1.95 &	411	 &  910	  & 2.52 & 0.05 & \\
		2126 &	8739&sm &	0.478 &	0.012 &	21	   & 0.70 & no	& --   & --   &  --  &  -- & 6547 & 47	 &   4.32 &	1.93 &	273	 &  947	  & 2.75 & 0.04 & \\
		2127 &	5734&sm &	0.482 &	0.015 &	43	   & 1.9  & no	& --   & --   &  --  &  -- & 6532 & 58	 &   4.32 &	1.91 &	236	 &  1554  & 2.8 & 0.07 & \\
		2129 &5021&	sm &	0.489 &	0.008 &	27	   & 1.2  & no	& --   & --   &  --  &  -- & 6501 & 31	 &   4.33 &	1.88 &	189	 &  1057  & 2.95 & 0.06 & \\
		2130 &	9741&sm &	0.508 &	0.011 &	31	   & 1.2  & no	& --   & --   & 2.25  & 0.17 & 6425 & 42	 &   4.34 &	1.80 &	327	 &  1196  & 2.78 & 0.04 & \\
		2131 &	6008&sm &	0.505 &	0.004$^*$ &<20& 0.5   & no	& --   & --   &  --  &  -- & 6437 & --	 &   4.34 &	1.81 &	231	 &  861	  & 2.6 & 0.06 & \\
		2132 &	10745&bm &	0.504 &	0.015 &	21	   & 1.0  & no	& --   & --   & 3.87  & 0.19 & 6440 & 57	 &   4.34 &	1.81 &	228	 &  993	  & 3.15 & 0.03 & \\
		2133 &	7733&sm &	0.498 &	0.013 &	32	   & 1.6  & no	& --   & --   & 2.11  & 0.13 & 6465 & 50	 &   4.33 &	1.84 &	199	 &  1152  & 2.8 & 0.06 & \\
		2134 &	7736&sm &	0.486 &	0.015 &	22	   & 1.0  & no	& --   & --   & 2.79  & 0.26 & 6515 & 58	 &   4.32 &	1.89 &	190	 &  991	  & 3.1 & 0.04 & \\
		2135 &	4760&sm &	0.510 &	0.011 &	21	   & 0.6  & no	& --   & --   & 2.65  & 0.16 & 6415 & 42	 &   4.34 &	1.79 &	334	 &  986	  & 3.13 & 0.03 & \\
		2137 &	4020&sm &	0.517 &	0.014 &	29	   & 1.1  & no	& --   & --   & 2.43  & 0.38 & 6387 & 53	 &   4.35 &	1.76 &	180	 &  1107  &	2.8 & 0.06 & \\
		2138 &	10747&sm &	0.529 &	0.010 &	22	   & 0.8  & no	& 3.25 & 0.08 & 3.01  & 0.18 & 6338 & 38	 &   4.36 &	1.71 &	210	 &  963	  & 2.95 & 0.04 & \\
		2139 &	9736&sm &	0.541 &	0.013 &	<20	   & 0.4  & no	& --   & --   & 3.28  & 0.74 & 6292 & 48	 &   4.37 &	1.66 &	180	 &  825	  & 3.1 & 0.04 & \\
		2140 &	7735&sm &	0.545 &	0.003 &	21 & 0.5	  & no	& --   & --   & 1.62  & 0.59 & 6276 & 11	 &   4.37 &	1.64 &	275	 &  893 & 2.97 & 0.03 & \\
		2141 &	9737& ulm &0.539 &	0.014&30&1.09 & no	& 1.80 & 0.02 & 2.52  & 0.48 & 6300 & 52	 &   4.37 &	1.67 &	319	 &  1111  &	3.24 & 0.03 & \\
		2145 &	11749&sm &	0.550 &	0.015 &	<20	   & 0.7  & no	& --   & --   & 2.27  & 0.36 & 6255 & 56	 &   4.37 &	1.62 &	160	 &  818	  & 3.18 & 0.04 & \\
		2146 &	12737&sm &	0.552 &	0.009 &	21	   & 0.7  & no	& --   & --   &  --  &  -- & 6249 & 33	 &   4.38 &	1.61 &	175	 &  922	  &  3.0 & 0.04 & \\
		2148 &	13745&sm &	0.573 &	0.009 &	<20	   & 0.7  & no	& --   & --   &  --  &  -- & 6168 & 33	 &   4.39 &	1.52 &	263	 &  900	  & 2.97 & 0.03 & \\
		2149 &	9732&sm &	0.569 &	0.011 &	<20	   & 0.6  & no	& --   & --   & 4.77  &  0.5 & 6184 & 40	 &   4.39 &	1.54 &	192	 &  866	  & 3.05 & 0.03 & \\
		2157 &	11741&sm &	0.598 &	0.012 &	<20	   & 0.2  & no	& --   & --   & 6.27  & 1.11 & 6070 & 43	 &   4.42 &	1.41 &	171	 &  901	  &  2.95 & 0.04 & \\
		2158 &	8740&sm &	0.610 &	0.010 &	<20	   & 0.2  & no	& --   & --   & 5.55  & 1.11 & 6026 & 35	 &   4.43 &	1.37 &	188	 &  903	  & 2.95 & 0.04 & \\
		2160 &	9022&sm &	0.625 &	0.004 &	<20	   & 0.1  & no	& --   & --   &  --  &  -- & 5969 & 14	 &   4.44 &	1.30 &	134	 &  883	  & 2.83 & 0.04 & \\
		2161 &	14747&sm &	0.608 &	0.015 &	<20	   & 0.1  & no	& 6.24 & 0.35 &  --  &  -- & 6034 & 53	 &   4.42 &	1.37 &	104	 &  907	  &  2.85 & 0.06 & \\
		2165 &	5011&sm &	0.635 &	0.010 &	<20	   & 0.2  & no	& --   & --   &  --  &  -- & 5931 & 35	 &   4.45 &	1.26 &	190	 &  832	  &  2.84 & 0.03 & \\
		2166 &	12759&sm &	0.644 &	0.002 &	<20	   & 0.3  & no	& --   & --   & 6.18  & 0.99 & 5896 & 7	 &   4.46 &	1.22 &	158	 &  849	  & 2.7 & 0.03 & \\
		2167 &	14736&sm &	0.654 &	0.004 &	<20	   & 0.3  & no	& 5.85 & 0.36 & 5.98  & 0.63 & 5862 & 14	 &   4.46 &	1.19 &	124	 &  800	  & 2.7 & 0.05 & \\
		2168 &	8015&sm &	0.630 &	0.005 &	<20	   & 0.3  & no	& 5.33 & 0.26 & 5.04  & 0.59 & 5951 & 17	 &   4.44 &	1.28 &	136	 &  813	  & 2.75 & 0.04 & \\
		2170 &	14025&sm &	0.651 &	0.004 &	<20	   & 0.4  & no	& 6.28 & 0.44 & 6.32  & 1.49 & 5873 & 14	 &   4.46 &	1.20 &	202	 &  879	  &  2.72 & 0.04 & \\
		2172 &	10760&sm &	0.661 &	0.008 &	<20	   & 0.3  & no	& --   & --   & 6.32  &  1.1 & 5837 & 27	 &   4.47 &	1.16 &	126	 &  827	  &  2.65 & 0.05 & \\
		2177 &	6006&sm &	0.694 &	0.003$^*$& <20& 0.2	  & no	& 5.95 & 0.32 &  --  &  -- & 5717 & --	 &   4.50 &	1.03 &	116	 &  742	  & 2.55 & 0.05 & \\
		2178 &	12742&sm &	0.685 &	0.023 &	<20	   & 0.16 & no	& --   & --   & 7.44  & 0.35 & 5748 & 77	 &   4.49 &	1.06 &	246	 &  835	  & 2.51 & 0.03 & \\
		2179 &	8013&sm &	0.695 &	0.011 &	<20     & 0.3 & no	& --   & --   & 6.95  & 1.12 & 5715 & 36	 &   4.50 &	1.02 &	98	 &  881	  &  2.4 & 0.07 & \\
		2184 &	14729& sm &	0.734 &	0.011 &	<20	   & 0.2  & no	& --   & --   &  --  &  -- & 5576 & 35	 &   4.53 &	0.87 &	94	 &  840	  & 2.1 & 0.07 & \\
		2185 &	10015&sm &	0.700 &	0.011 &	<20	   & 0.3  & no	& --   & --   & 7.31  & 1.32 & 5697 & 36	 &   4.50 &	1.01 &	118	 &  834	  &  2.35 & 0.05 & \\
		2187 &	10740&sm &	0.724 &	0.011 &	<20	   & 0.2  & no	& 6.21 & 0.36 & 6.16  & 0.74 & 5611 & 35	 &   4.52 &	0.91 &	160	 &  831	  &  2.5 & 0.04 & \\
		2188 &	11735&sm &	0.721 &	0.004 &	<20	   & 0.2  & no	& 7.83 & 1.73 & 6.13  & 1.18 & 5622 & 13	 &   4.52 &	0.92 &	108	 &  879	  & 2.3 & 0.04 & \\
		2193 &	12026&sm &	0.726 &	0.010 &	<20	   & 0.1  & no	& --   & --   &  --  & 0.15 & 5604 & 32	 &   4.52 &	0.90 &	117	 &  859	  & 2.15 & 0.05 & \\
		2195 &	15734&sm &	0.747 &	0.005 &	<20	   & 0.2  & no	& --   & --   & 6.61  & 0.92 & 5533 & 16	 &   4.54 &	0.83 &	122	 &  800	  & 2.12 & 0.04 & \\
		2197 &	12730& sm &	0.735 &	0.005 &	<20	   & 0.2  & no	& 7.19 & 0.38 &  --  &  -- & 5575 & 16	 &   4.53 &	0.87 &	95	 &  830	  & 2.2 & 0.07 & \\
		2202 &	--&sm &	0.765 &	0.007 &	<20	   & 0.2	  & no	& --   & --   &  --  &  -- & 5471 & 22	 &   4.55 &	0.80 &	122	 &  807	  & 2.1 & 0.04 & \\
		2203 &	16742&sm &	0.770 &	0.022 &	<20	   & 0.14 & no	& --   & --   & 7.31  & 1.19 & 5456 & 68	 &   4.55 &	0.80 &	160	 &  811	  & 1.88 & 0.07 & \\
		2204 &	14730&sm &	0.770 &	0.007 &	<20	   & 0.2  & no	& 7.35 & 0.44 & 7.34  & 0.71 & 5456 & 22	 &   4.55 &	0.80 &	115	 &  830	  & 1.8 & 0.05 & \\
		2207 &	--&sm &	0.781 &	0.006 &	<20& 0.2          & no	& --   & --   &  --  &  -- & 5418 & 18	 &   4.56 &	0.80 &	109	 &  836	  & 2.1 & 0.05 & \\
		2212 &	7006&sm &	0.782 &	0.005$^*$& <20 & 0.3  & no	& --   & --   & 6.92  & 0.21 & 5415 & --	 &   4.56 &	0.80 &	89	 &  867	  &  2.0 & 0.08 & \\
		2213 &	23737&sm &	0.780 &	0.009 &	<20	   & 0.1  & no	& 7.90 & 0.80 & 7.13  & 1.06 & 5421 & 28	 &   4.56 &	0.80 &	97	 &  842	  & 1.8 & 0.08 & \\
		2217 &	16028& sm &	0.800 &	0.007 &	<20	   & 0.2  & no	& 7.47 & 0.58 &  --  &  -- & 5356 & 21	 &   4.57 &	0.80 &	117	 &  824	  & 1.9 & 0.07 & \\
		2221 &	26743&sm &	0.814 &	0.005 &	<20	   & 0.2  & no	& --   & --   & 7.28  & 0.74 & 5312 & 15	 &   4.58 &	0.80 &	97	 &  837	  & 1.45 & 0.08 & \\
		2222 &	14746&sm &	0.806 &	0.011 &	<20	   & 0.2  & no	& --   & --   &  --  &  -- & 5335 & 33	 &   4.58 &	0.80 &	133	 &  835	  & 1.95 & 0.07 & \\
		2223 &	20742&sm &	0.808 &	0.011 &	<20	   & 0.2  & no	& 8.31 & 0.57 & 8.21  & 1.27 & 5332 & 33	 &   4.58 &	0.80 &	100	 &  776	  & 1.55 & 0.08 & \\
		2224 &	15016&sm &	0.825 &	0.008 &	<20	   & 0.2  & no	& --   & --   &  --  &  -- & 5275 & 24	 &   4.59 &	0.80 &	100	 &  834	  & 1.40 & 0.08 & \\
		2227 &	15731&sm &	0.821 &	0.006 &	<20	   & 0.2  & no	& --   & --   &  --  &  -- & 5290 & 18	 &   4.58 &	0.80 &	114	 &  814	  & 1.35 & 0.07 & \\
		2228 &	8010&sm &	0.831 &	0.005 &	<20	 & 2.4	  & no	& 7.96 & 0.64 & 7.42  & 0.16 & 5258 & 15	 &   4.59 &	0.80 &	137	 &  1012  & 1.60 & 0.07 & \\
		2231 &	3003&sm &	0.742 &	0.004$^*$& <20& 0.2	  & no	& --   & --   &  --  &  -- & 5550 & --	 &   4.53 &	0.85 &	120	 &  902	  & 1.95 & 0.07 & \\
		2232 &	16016&sm &	0.861 &	0.005 &	<20	   & 0.13 & no	& 8.23 & 0.51 &  8.4  & 0.17 & 5163 & 14	 &   4.61 &	0.80 &	163	 &  814	  & <1.0  & 0.06 & \\
		2233 &	23750& sm &	0.847 &	0.011 &	<20	   & 0.2  & no	& --   & --   &  --  &  -- & 5207 & 32	 &   4.60 &	0.80 &	86	 &  815	  &  1.80 & 0.08 & \\
		2234 &	24744&sm &	0.863 &	0.010 &	<20	   & 0.2  & no	& --   & --   &  --  &  -- & 5156 & 28	 &   4.61 &	0.80 &	100	 &  804	  & 1.25 & 0.08 & \\
		2235 &	17023&sm &	0.861 &	0.005 &	<20	   & 0.2  & no	& 8.75 & 0.79 & 7.40  & 0.68 & 5164 & 14	 &   4.61 &	0.80 &	174	 &  818	  & 1.65 & 0.07 & \\
		2237 &	26744&sm &	0.859 &	0.007 &	<20	   & 0.2  & no	& 8.63 & 0.61 &  --  &  -- & 5169 & 20	 &   4.60 &	0.80 &	83	 &  795	  & 1.20 & 0.13 & \\
		2238 &	26747&sm &	0.913 &	0.012 &	<20	   & 0.2  & no	& --   & --   &  --  &  -- & 5006 & 32	 &   4.63 &	0.80 &	131 &  825	  & 1.40 & 0.09 & \\
		2239 &	24740&sm &	0.900 &	0.006 &	<20	   & 0.2  & no	& --   & --   & 7.76  & 0.94 & 5044 & 16	 &   4.62 &	0.80 &	82	 &  802	  &  <1.45  & -- & \\
		2241 &	26740&sm &	0.894 &	0.008 &	<20	   & 0.2  & no	& 7.52 & 0.45 & 7.04  & 1.16 & 5063 & 22	 &   4.62 &	0.80 &	116	 &  830	  &  1.62 & 0.07 & \\
		2242 &	26738&sm &	0.922 &	0.009 &	<20	   & 0.3  & no	& 8.03 & 0.55 &  --  &  -- & 4979 & 24	 &   4.63 &	0.80 &	105	 &  817	  & 1.45 & 0.07 & \\
		2244 &	21020&sm &	0.916 &	0.005 &	<20	   & 0.2  & no	& 7.82 & 0.54 & 7.62  & 1.05 & 4998 & 13	 &   4.63 &	0.80 &	94	 &  825	  & 1.50 & 0.07& \\
		2245 &	27740&sm &	0.930 &	0.003 &	<20	   & 1.1  & no	& 7.86 & 0.47 & 8.07  &  1.1 & 4957 & 8	 &   4.64 &	0.80 &	109	 &  840	  & 1.55 & 0.06 & \\
		2246 &	18018&sm &	0.929 &	0.005 &	<20	   & 0.3  & no	& --   & --   &  --  &  -- & 4960 & 13	 &   4.64 &	0.80 &	99	 &  821	  & 0.85 & 0.06 & \\
		2247 &	21023&sm &	0.933 &	0.006 &	<20	   & 0.3  & no	& 9.24 & 0.75 &  8.5  & 1.49 & 4948 & 16	 &   4.64 &	0.80 &	97	 &  888	  & 1.10 & 0.08 & \\
		2249 &	5004&sm &	0.929 &	0.005 &	<20	   & 0.3  & no	& 7.52 & 0.50 &  --  &  -- & 4958 & 13	 &   4.64 &	0.80 &	84	 &  806	  & 1.45 & 0.08 & \\
		2250 &	13011&sm &	0.929 &	0.007 &	<20	   & 0.2  & no	& --   & --   &  --  &  -- & 4959 & 19	 &   4.64 &	0.80 &	107	 &  855	  &  <0.90  & --& \\
		2251 &	--&sm &	0.943 &	0.004$^*$	& <20  & 0.2  & no	& --   & --   &  --  & 0.08 & 4919 & --	 &   4.64 &	0.80 &	69	 &  836	  &  <0.90  & --& \\
		2252 &	48754&sm &	0.935 &	0.003 &	<20	   & 0.2  & no	& 8.75 & 0.68 & 8.49  & 0.45 & 4943 & 8	 &   4.64 &	0.80 &	121	 &  811	  & 1.10 & 0.08 & \\
		2254 &	6004& sm &	0.936 &	0.004$^*$& <20 & 0.2  & no	& --   & --   &  --  &  -- & 4939 & --	 &   4.64 &	0.80 &	75	 &  859	  & <0.80 & --& \\
		2256 &	10012& sm &	0.957 &	0.007 &	<20	   & 0.2  & no	& 9.27 & 0.68 &  --  &  -- & 4880 & 18	 &   4.65 &	0.80 &	105	 &  823	  & <0.50  & --& \\
		2257 &	20016&sm &	0.946 &	0.008 &	<20	   & 0.2  & no	& 9.55 & 0.90 &  8.7  & 1.13 & 4910 & 21	 &   4.64 &	0.80 &	86	 &  841	  & <0.65  & --& \\
		2262 &	41752&sm &	0.955 &	0.006 &	<20	   & 0.2  & no	& 9.58 & 0.62 & 8.37  & 1.42 & 4885 & 15	 &   4.65 &	0.80 &	70	 &  942	  & 0.80 & 0.14 & \\
		2263 &	21016&sm &	0.964 &	0.009 &	<20	   & 0.2  & no	& --   & --   & 8.03  & 1.27 & 4861 & 23	 &   4.65 &	0.80 &	75	 &  834	  & 0.70 & 0.13 & \\
		2266 &	12007&sm &	0.975 &	0.005 &	<20	   & 0.2  & no	& --   & --   &  --  &  -- & 4829 & 13	 &   4.66 &	0.80 &	67	 &  787	  & <0.65 & -- & \\
		2267 &	30753&sm &	0.992 &	0.002 &	<20	   & 0.5  & no	& --   & --   & 8.39  & 1.49 & 4782 & 5	 &   4.66 &	0.80 &	89	 &  870	  & <0.60 & --& \\
		2269 &	41749&sm &	0.971 &	0.007 &	<20	   & 0.2  & no	& --   & --   &  --  &  -- & 4841 & 18	 &   4.65 &	0.80 &	91	 &  822	  & 1.15 & 0.09 & \\
		2273 &	35742&sm &	0.991 &	0.005 &	<20	   & 0.2  & no	& --   & --   &  --  &  -- & 4785 & 12	 &   4.66 &	0.80 &	89	 &  675	  & <0.60  & -- & \\
		2274 &	41737&sm &	0.998 &	0.010 &	<20	   & 0.19 & no	& --   & --   & 7.51  & 0.85 & 4764 & 25	 &   4.66 &	0.80 &	159	 &  826	  & 0.73 & 0.07 & \\
		2276 &	18019&sm &	0.990 &	0.006 &	<20	   & 0.2  & no	& --   & --   &  8.5  &  0.9 & 4786 & 15	 &   4.66 &	0.80 &	104	 &  817	  & <0.60 & -- & \\
		2281 &	28022&sm &	1.014 &	0.008 &	<20	   & 0.2  & no	& --   & --   &  --  &  -- & 4721 & 19	 &   4.67 &	0.80 &	139	 &  805	  & 0.58 & 0.14 & \\
		2283 &	47756&sm &	1.008 &	0.003 &	<20	   & 0.3  & no	& --   & --   &  --  &  -- & 4738 & 7	 &   4.67 &	0.80 &	67	 &  838	  & <0.65  & -- & \\
		2285 &	20019&sm &	1.037 &	0.008 &	<20	   & 0.2  & no	& 8.83 & 0.72 &  8.5  & 1.03 & 4661 & 19	 &   4.68 &	0.80 &	97	 &  848	  & 0.95 & 0.08 & \\
		2286 &	33734&sm &	1.016 &	0.014 &	<20	   & 0.2  & no	& --   & --   & 6.32  & 0.81 & 4719 & 34	 &   4.67 &	0.80 &	97	 &  822	  & 0.95 & 0.08 & \\
		2288 &	16010&sm &	1.038 &	0.020 &	<20	   & 0.2  & no	& --   & --   & 8.25  & 1.71 & 4659 & 47	 &   4.68 &	0.80 &	89	 &  663	  & 0.60 & 0.08 & \\
		2289 &	17017&sm &	1.043 &	0.008 &	<20	   & 0.2  & no	& --   & --   &  --  &  -- & 4647 & 19	 &   4.68 &	0.80 &	87	 &  833	  & <0.40  & -- & \\
		2291 &	11006&sm &	1.018 &	0.004$^*$& <20& 0.2	  & no	& --   & --   &  --  &  -- & 4712 & --	 &   4.67 &	0.80 &	83	 &  833	  & 0.90 & 0.09 & \\
		2294 &	41740&sm &	1.069 &	0.008 &	<20   & 0.2	  & no	& --   & --   &  --  &  -- & 4579 & 18	 &   4.69 &	0.80 &	82	 &  870	  & 0.50 & 0.10 & \\
		2297 &	42747&sm &	1.071 &	0.022 &	<20	   & 0.3  & no	& --   & --   &  --  &  -- & 4576 & 50	 &   4.69 &	0.80 &	73	 &  838	  & <0.35 & --& \\
		2298 &	41738&sm &	1.110 &	0.010 &	<20	   & 0.2  & no	& --   & --   &  --  &  -- & 4481 & 22	 &   4.70 &	0.80 &	86	 &  797	  & <0.15  & --& \\
		2302 &	33022&sm &	1.106 &	0.012 &	<20	   & 0.2  & no	& --   & --   &  --  &  -- & 4489 & 26	 &   4.70 &	0.80 &	75	 &  889	  & <0.25 & -- & \\
		2303 &	27016&sm &	1.105 &	0.011 &	<20	   & 0.3  & no	& --   & --   &  --  &  -- & 4491 & 24	 &   4.70 &	0.80 &	107	 &  661	  & <0.0 & -- & \\
		2305 &	29016&sm &	1.109 &	0.007 &	<20	   & 0.2  & no	& --   & --   &  --  &  -- & 4482 & 15	 &   4.70 &	0.80 &	65	 &  819	  & 0.80 & 0.11 & \\
		2308 &	23015&sm &	1.112 &	0.007 &	<20	   & 0.2  & no	& 4.81 & 0.24 &  --  &  -- & 4476 & 15	 &   4.71 &	0.80 &	94	 &  856	  & 0.50 & 0.08 & \\
		2309 &	30026&sm &	1.115 &	0.007 &	<20	   & 0.2  & no	&13.31 & 1.58 &  --  &  -- & 4468 & 15	 &   4.71 &	0.80 &	83	 &  876	  &  0.40 & 0.16 & \\
		2311 &	50748&sm &	1.137 &	0.012 &	<20	   & 0.2  & no	& --   & --   &  --  &  -- & 4416 & 25	 &   4.72 &	0.80 &	75	 &  750	  &  <0.20  & --& \\
		2318 &	23024& slm &1.176 &0.027&<20&0.2 & no	& --   & --   &  --  &  -- & 4327 & 53	 &   4.73 &	0.80 &	85	 &  685	  & <0.10  & --& \\
		2319 &	-- & sm &	1.139 &	0.013 &	<20	   & 0.2  & no	& --   & --   & 7.08  & 0.18 & 4412 & 27	 &   4.72 &	0.80 &	71	 &  785	  &  <0.20  & --& \\
		2323 &	65752 &sm &	1.156 &	0.014 &	<20  & 0.3	  & no	& --   & --   & 8.57  & 0.18 & 4373 & 28	 &   4.72 &	0.80 &	68	 &  652	  &  <0.25 & -- & \\
		2324 &	32016& sm &	1.183 &	0.014 &	<20	   & 0.3  & no	& --   & --   & 2.14  &  0.5 & 4312 & 27	 &   4.73 &	0.80 &	63	 &  912	  &  <0.15  & --& \\
		2326 &	60748&sm &	1.166 &	0.016 &	<20	   & 0.3  & no	& 5.81 & 0.30 &  --  & 0.52 & 4350 & 32	 &   4.73 &	0.80 &	87	 &  833	  & 0.10 & 0.18 & \\
		2327 &	13012&sm &	1.160 &	0.012 &	<20   & 0.3	  & no	& --   & --   &  --  &  -- & 4364 & 24	 &   4.72 &	0.80 &	58	 &  641	  & <0.20 & -- & \\
		2329 &	61748&sm &	1.162 &	0.012 &	<20   & 0.3	  & no	& --   & --   &  --  &  -- & 4359 & 24	 &   4.72 &	0.80 &	71	 &  904	  & 0.15 & 0.19 & \\
		2331 &	24017&sm &	1.166 &	0.010 &	<20	   & 0.3  & no	&10.22 & 0.84 &  --  &  -- & 4350 & 20	 &   4.73 &	0.80 &	86	 &  806	  & <0.10 & -- & \\
		2337 &	55739&bm &	1.216 &	0.017 &	<20	   & 0.6  & no	& --   & --   &  --  &  -- & 4240 & 32	 &   4.75 &	0.80 &	86	 &  940	  &  <-0.1 & --& \\
		2338 &	62737&sm &	1.204 &	0.004 &	<20	   & 0.3  & no	& 6.79 & 0.44 &  --  &  -- & 4267 & 8	 &   4.74 &	0.80 &	77	 &  822	  & <-0.05 & --& \\
		2341 &	74756&sm &	1.210 &	0.013 &	<20	   & 0.3  & no	& --   & --   & 3.39  & 0.52 & 4253 & 24	 &   4.74 &	0.80 &	94	 &  792	  & <-0.05 & --& \\
		2344 &	32035&sm &	1.227 &	0.026 &	<20	   & 0.3  & no	& --   & --   &  --  &  -- & 4217 & 48	 &   4.75 &	0.80 &	62	 &  861	  &  <0.2  & --& \\
		2348 &	47736&sm &	1.213 &	0.016 &	<20	 & 0.8	  & no	& --   & --   &  --  &  -- & 4247 & 30	 &   4.75 &	0.80 &	74	 &  846	  &  <0.1  & --& \\
		2349 &	32024&sm &	1.248 &	0.016 &	<20	   & 0.3  & no	& --   & --   & 9.33  & 0.11 & 4175 & 28	 &   4.76 &	0.80 &	56	 &  832	  & <0.1  & --& \\
		2351 &	62747&sm &	1.239 &	0.005 &	<20	 & 0.4	  & no	& --   & --   &  --  &  -- & 4194 & 9	 &   4.76 &	0.80 &	61	 &  801	  & <0.0  & --& \\
		2352 &	53759&sm &	1.264 &	0.010 &	<20 & 0.4	  & no	& --   & --   &  --  &  -- & 4143 & 17	 &   4.76 &	0.80 &	72	 &  713	  & <0.0  & --& \\
		2354 &	41028&sm &	1.260 &	0.012 &	<20 & 0.3	  & no	& --   & --   &  --  &  -- & 4150 & 21	 &   4.76 &	0.80 &	61	 &  785	  &  <0.0  & --& \\
		2356 &	33024&sm &	1.257 &	0.015 &	<20	   & 0.4  & no	& --   & --   &  --  &  -- & 4156 & 26	 &   4.76 &	0.80 &	55	 &  759	  & <0.05 & -- & \\
		2357 &	27015&sm &	1.260 &	0.013 &	<20 & 0.4	  & no	& --   & --   &  --  &  -- & 4150 & 23	 &   4.76 &	0.80 &	61	 &  942	  &  <0.05  & --& \\
		2359 &	45027&sm &	1.273 &	0.014 &	<20   & 0.5	  & no	& --   & --   &  --  &  -- & 4125 & 24	 &   4.77 &	0.80 &	69	 &  881	  &  <-0.1  & --& \\
		2370 &	62759&sm &	1.309 &	0.024 &	<20	   & 0.4  & no	& --   & --   & 6.72  & 0.09 & 4056 & 38	 &   4.78 &	0.80 &	75	 &  741	  & <-0.1 & -- & \\
		2371 &	87755&sm &	1.307 &	0.014 &	<20	   & 0.4  & no	& --   & --   &  --  &  -- & 4060 & 23	 &   4.78 &	0.80 &	78	 &  1028  &	 <0.0  & --& \\
		3005 &	4023&bm &	0.243 &	0.023 &	56     & 10.4 & no	& --   & --   &  --  &  -- & 7590 & 107&	4.25 &	2.85 &	603	 &  2983  &	<3.3 & --& \\
		3008 &	1759&bm &	0.189 &	0.038 &	67	   & 5.7  & no	& --   & --   &  --  &  -- & 7856 & 179&	4.24 &	3.08 &	510	 &  2530  &	  <3.2 & -- & \\
		3010 &	2731&sm &	0.164 &	0.197 &	53	   & 3.3  & no	& --   & --   &  --  &  -- & 7981 & 939&	4.21 &	3.21 &	594	 &  1873  &	 <2.9 & --& \\
		3016 &	4013&bm &	0.467 &	0.013 &	30	   & 1.0  & no	& --   & --   & 1.21  & 0.11 & 6577 & 51	 &   4.31 &	1.96 &	330	 &  1129  &	2.4 & 0.04 & \\
		3018 &	4014&bm &	0.507 &	0.013 &	49	   & 8.5  & no	& --   & --   &  --  &  -- & 6415 & 50	 &   4.34 &	1.79 &	168	 &  2096  &	 <2.6  & --& \\
		3019 &	3735& um &	0.421 &0.020 &<20&0.7 & no	& --   & --   &  --  &  -- & 6777 & 79	 &   4.28 &	2.16 &	297	 &  920	  &  2.85 & 0.05 & \\
		3021 &	--&sm &	0.459 &	0.026 &	49	   & 5.5	  & no	& --   & --   &  --  &  -- & 6614 & 105 &	4.30 &	2.00 &	277	 &  1752  &	 2.6 & 0.11 & \\
		3022 &	9018&bm &	0.664 &	0.018 &	28	   & 1.4  & no	& --   & --   & 1.91  & 0.65 & 5801 & 60	 &   4.47 &	1.13 &	211	 &  1082  &	 2.15 & 0.05 & \\
		3024 &	4021& um &	0.585&0.007&34& 1.9	  & no	& --   & --   &  --  &  -- & 6101 & 24	 &   4.41 &	1.45 &	283	 &  1271  &	  3.0 & 0.03 & \\
		3026 &	9729&bm &	0.602 &	0.010 &	71	   & 4.6  & no	& --   & --   & 2.18  & 0.29 & 6035 & 37	 &   4.42 &	1.38 &	281	 &  2850  &	 2.7 & 0.06 & \\
		3029 &	5744&sm &	0.496 &	0.019 &	<20	   & 0.6  & no	& --   & --   &  --  &  -- & 6458 & 72	 &   4.33 &	1.84 &	154	 &  841	  & 3.1 & 0.04 & \\
		3030 &	7021& um &	0.543 &	0.014 &29&1.6 & no	& 2.86 & 0.10 & 3.13  & 0.34 & 6268 & 54	 &   4.37 &	1.64 &	263	 &  1096  &	2.3 & 0.04 & \\
		3031 &	11747& um &0.654 &	0.012 &21&0.4 & no  & --   & --   & 5.94  & 1.04 & 5839 & 40	 &   4.46 &	1.17 &	234	 &  946	  &  2.75 & 0.04 & \\
		3033 &	4010&bm &	0.647 &	0.003 &	45	   & 1.6  & no	& --   & --   &  --  &  -- & 5865 & 9	 &   4.46 &	1.20 &	195	 &  1672  &	2.45 & 0.04 & \\
		3035 &	5020& ulm &0.563 &	0.031&<20&0.9 & no	& --   & --   &  --  &  -- & 6188 & 114 &	4.39 &	1.55 &	262	 &  912	  &  3.15 & 0.05 & \\
		3036 &	10737&sm &	0.567 &	0.012 &	<20	 & 0.6	  & no	& --   & --   & 4.47  & 0.61 & 6172 & 45	 &   4.39 &	1.53 &	204	 &  859	  & 2.9 & 0.04 & \\
		3037 &	11752&sm &	0.554 &	0.016 &	<20  & 0.8	  & no	& --   & --   & 5.01  & 0.69 & 6223 & 58	 &   4.38 &	1.59 &	186	 &  879	  & 3.15 & 0.04 & \\
		3039 &	10736&bm &	0.674 &	0.008 &	24	   & 1.7  & no	& --   & --   & 7.04  & 0.88 & 5765 & 27	 &   4.48 &	1.09 &	307	 &  1034  & --	& -- & sb2 \\
		3040 &	10756&sm &	0.581 &	0.013 &	<20	   & 0.8  & no	& --   & --   & 4.07  & 0.51 & 6118 & 45	 &   4.40 &	1.47 &	235	 &  892	& 2.9 & 0.03 & \\
		3041 &	7744&bm &	0.702 &	0.017 &	32	   & 1.6  & no	& 7.72 & 0.56 & 7.82  & 1.18 & 5667 & 55	 &   4.50 &	0.98 &	338	 &  1300  &	2.4 & 0.03 & \\
		3042 &	8735& um &	0.579&0.016&<20& 0.6  & no	& 6.12 & 0.41 & 5.94  & 0.48 & 6126 & 60	 &   4.40 &	1.48 &	201	 &  857	  & 3.05 & 0.04 & \\
		3043 &	--&sm &	0.621 &	0.008 &	<20	   & 0.2	  & no	& --   & --   & 2.79  & 0.33 & 5965 & 26	 &   4.44 &	1.31 &	174	 &  878	  &  2.87 & 0.04 & \\
		3044 &	7015&bm &	0.756 &	0.029 &	37	   & 2.5  & no	& --   & --   & 2.18  & 0.64 & 5479 & 92	 &   4.54 &	0.80 &	340	 &  1386  & 2.1 & 0.04 & \\
		3045 &	5010&bm &	0.754 &	0.026 &	52	   & 5.1  & no	& --   & --   & 1.91  & 0.22 & 5483 & 81	 &   4.54 &	0.80 &	249	 &  1896  & 2.1 & 0.04 & \\
		3046 &	6013&bm &	0.899 &	0.018 &	50	   & 1.9  & no	& --   & --   & 8.48  & 0.74 & 5018 & 49	 &   4.62 &	0.80 &	239	 &  1824  &	 <1.4  & -- & \\
		3047 &	10738&sm &	0.632 &	0.013 &	<20	   & 0.3  & no	& 6.42 & 0.35 & 6.63  & 1.17 & 5922 & 46	 &   4.45 &	1.26 &	231	 &  851	  & 2.8 & 0.03 & \\
		3049 &	6011&sm &	0.665 &	0.038 &	<20	   & 0.6  & no	& --   & --   &  --  &  -- & 5799 & 130 &	4.47 &	1.12 &	199	 &  906	  &  2.62 & 0.03 & \\
		3050 &	18743&sm &	0.670 &	0.031 &	<20	   & 0.2  & no	& --   & --   & 6.68  & 1.13 & 5781 & 105 &	4.48 &	1.10 &	133	 &  840	  &  2.7 & 0.05 & \\
		3052 &	9015&bm &	0.837 &	0.038 &	<20	   & 0.3  & no	& --   & --   & 3.28  & 0.24 & 5209 & 112 &	4.59 &	0.80 &	181	 &  906	  & 1.7 & 0.05 & \\
		3053 &	13734&sm &	0.732 &	0.018 &	<20	   & 0.2  & no	& 7.50 & 0.52 &  --  &  -- & 5559 & 58	 &   4.53 &	0.86 &	204	 &  885	  & 2.05 & 0.03 & \\
		3054 &	21755&sm &	0.730 &	0.008 &	<20	   & 0.2  & no	& --   & --   & 4.43  & 1.18 & 5566 & 26	 &   4.52 &	0.87 &	82	 &  863	  & 2.2 & 0.07 & \\
		3055 &	17734&sm &	0.820 &	0.030 &	<20	   & 0.2  & no	& 8.57 & 0.72 &  --  &  -- & 5264 & 89	 &   4.58 &	0.80 &	114	 &  842	  & 1.7& 0.07 & \\
		3056 &	12016& um &0.840&0.024&<20& 0.3	  & no	& --   & --   & 3.46  & 0.23 & 5200 & 69	 &   4.59 &	0.80 &	256	 &  903	  & 1.55 & 0.04 & \\
		3058 &	17758&sm &	0.797 &	0.009 &	<20   & 0.3	  & no	& --   & --   & 7.55  & 1.28 & 5339 & 26	 &   4.57 &	0.80 &	135	 &  876	  &  1.6 & 0.06 & \\
		3059 &	15740&sm &	0.805 &	0.009 &	<20	   & 0.2  & no	& 7.59 & 0.48 & 7.05  & 0.41 & 5315 & 26	 &   4.57 &	0.80 &	145	 &  851	  &  1.95 & 0.06 & \\
		3060 &	18732&sm &	0.822 &	0.023 &	<20	   & 0.2  & no	& --   & --   &  --  &  -- & 5259 & 67	 &   4.58 &	0.80 &	108	 &  865	  &  1.85 & 0.08 & \\
		3063 &	21027&sm &	0.865 &	0.009 &	<20	   & 0.2  & no	& --   & --   &  --  &  -- & 5121 & 24	 &   4.61 &	0.80 &	101	 &  858	  & 1.15 & 0.07 & \\
		3066 &	24760& um &0.902&0.009&<20 & 0.2 & no	& --   & --   &  3.8  & 0.20 & 5007 & 25	 &   4.63 &	0.80 &	58	 &  812	  &  <1.3  & -- & \\
		3067 &	28739& um &0.953&0.014&<20 & 0.4 & no	& --   & --   & 3.69  & 0.28 & 4860 & 37	 &   4.65 &	0.80 &	70	 &  872	  & 1.3 & 0.11 & \\
		3069 &	34751 &sm &	0.980 &	0.008 &	<20   & 0.3	  & no	& --   & --   &  --  &  -- & 4783 & 19	 &   4.66 &	0.80 &	87	 &  910	  & <0.5  & --& \\
		3070 &	28732& um &0.993&0.009&<20&0.3	  & no	& --   & --   & 7.19  & 0.15 & 4745 & 23	 &   4.66 &	0.80 &	95	 &  895	  & <0.5 & --& \\
		\hline
		\enddata
		\tablecomments{1. WIYN Open Cluster Id (WOCS Id) from the photometry paper (in preparation).
			2. Averaged $B-V$ ($(B-V)_{\rm eff}$) and standard deviation by using all 10 possible color combinations from $UBVRI$. $^*$ the star has only one measurement of just one color, so we show the SNR-based $\sigma$ from DAOPHOT. 
			3. Rotational velocity ({\it v} sin{\it i}) and errors in km s$^{-1}$. Column 8 indicates whether the {\it v} sin {\it i} is from $H_{\alpha}$ line fitting.
			4. Rotational period and error in days, from \citet{2015AA...583A..73B}.
			5. Rotational period and error in days, derived from TESS light curves.
			6. Stellar atmospheric parameters including $T_{\rm eff}$, error on $T_{\rm eff}$, log {\it g}, and microturbulence ($V_{\rm t}$).
			7. Signal-to-noise ratio (SNR) of the combined spectra, full width at half maximum (FWHM).
			8. Error due to equivalent width and error due to $T_{\rm eff}$, added in quadrature.
			* means either multiplicity or membership has been changed from Paper I.
		}
	\end{deluxetable*}
\end{longrotatetable}

\subsection{Stellar Parameters}

As in Paper 1, we derive average $B-V$ ($(B-V)_{\rm eff}$) using all ten possible color combinations of $UBVRI$ from our own photometry, and calculate errors based on the standard deviation of the mean ($\sigma_{\mu}$, Columns 4 and 5). We followed the same procedures as in Paper 1 to derive $T_{\rm eff}$, log {\it g}, and $V_{\rm t}$, except we assumed [Fe/H] = -0.063 dex (instead of -0.05 dex) so the derived log {\it g} and $V_{\rm t}$ are very slightly different. Table \ref{tab:candidates} shows $T_{\rm eff}$, error on $T_{\rm eff}$ (propagated from $(B-V)_{\rm eff}$), log {\it g}, and $V_{\rm t}$ in columns 13, 14, 15, 16, respectively.  The first three rows include a subgiant and two giants whose location on the CMD are shown in Figures 6 and 8 of Paper 1. Our dwarfs span a very large $T_{\rm eff}$ range of 8575 -- 4056 K from early A to late K stars.

\section{Li Abundances}    \label{sec:li}

We employ spectrum synthesis near the Li I 6707.8 \AA\ feature to derive A(Li), and employ the refined line-list near Li from \citet{2022MNRAS.513.5387S}.  This line-list produces more accurate A(Li) in cooler stars where the neighboring Fe I 6707.43 \AA\ line grows stronger. Synthesis also produces more reliable A(Li) in rapid rotators where additional features may also blend with the Li line.  See \citet{2022MNRAS.513.5387S} for additional discussion about the line-list. 

\begin{figure}
	\centering
	\includegraphics[width=0.5\textwidth]{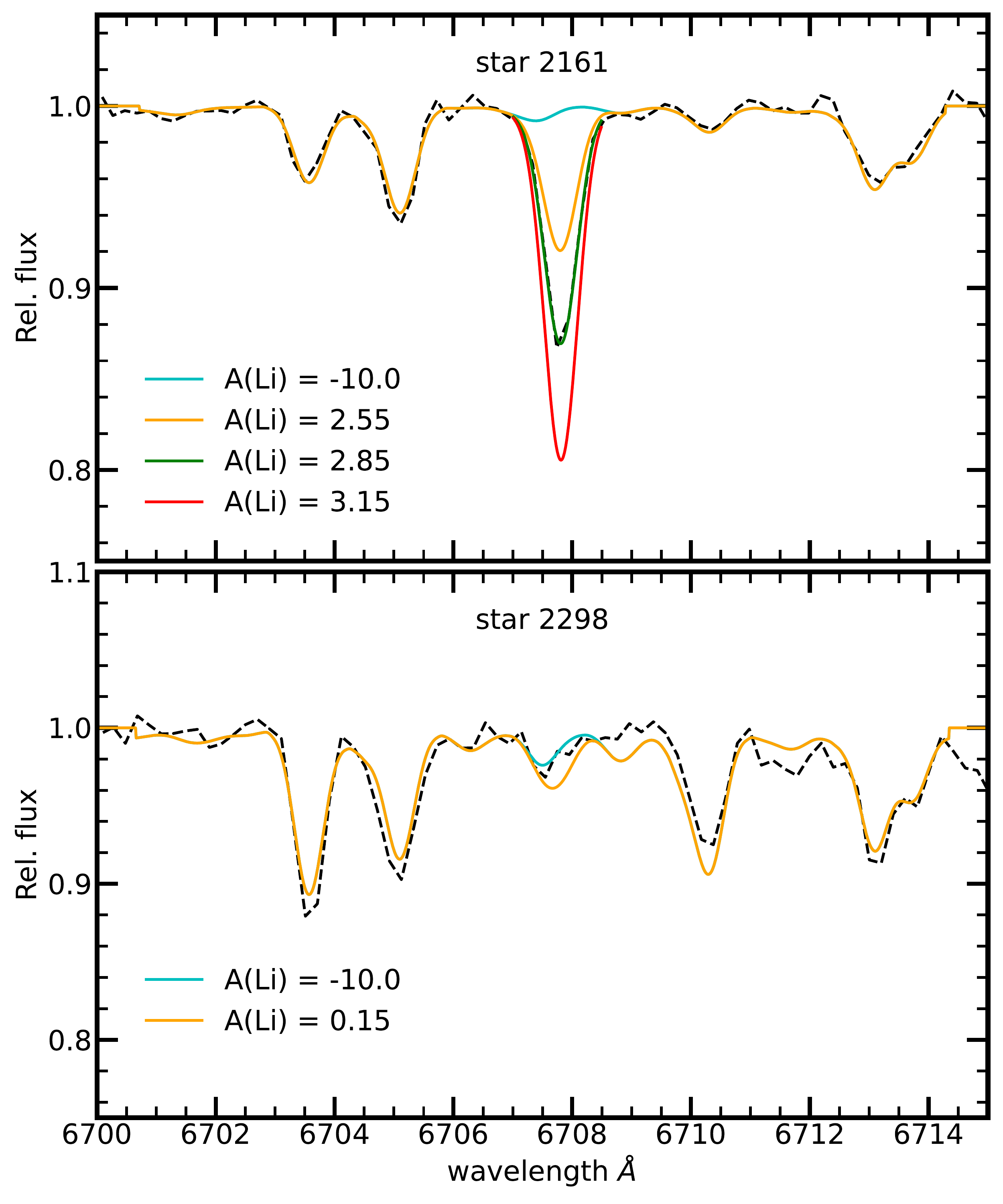}
	\caption{The top panel shows observed and synthetic spectrum for star 2161 ($T_{\rm eff}$ = 6034 K). The cyan line shows synthesis for no Li (A(Li) = -10.0 dex). The green line shows the best fit A(Li), and the orange and red lines show the A(Li) that are twice/half than the best fit A(Li). The bottom panel shows the 3$\sigma$ upper limit A(Li) for star 2298 ($T_{\rm eff}$ = 4481 K). The cyan line corresponds to synthesis of no Li, and the orange line shows an upper limit A(Li) of 0.15 dex.}
	\label{fig:star2161_2298}
\end{figure}

We use the relationship (equation \ref{eqn1}) taken from \citet{1993ApJ...414..740D} to separate Li detections from 3$\sigma$ upper limits.  
\begin{equation} \label{eqn1} 
	3 \sigma\ EW =3 \times 1.503 \times \frac{\sqrt{FWHM \times pixel\ scale }}{SNR}
\end{equation}
For detections, we use the {\it synth} task in MOOG to generate synthetic spectra in the Li region from 6700 to 6715 \AA\ . The top panel of Figure \ref{fig:star2161_2298} shows an example synthesis for star 2161, where the best fit A(Li) = 2.85 dex (green line).  For upper limits we report the A(Li) that corresponds to an equivalent width that equals the 3$\sigma$ value. These may be slightly high in cases where the Fe I 6707.43 \AA\ line becomes more significant (K dwarfs) or where rapid rotation blends in additional features (A dwarfs). In a few cases, we report a slightly higher upper limit as suggested by visual inspection of the spectra.  Column 19 shows the A(Li). The bottom panel of Figure \ref{fig:star2161_2298} shows an example synthesis for star 2298 as a 3$\sigma$ upper limit A(Li), where the star does not show a convincing detection of Li. Table \ref{tab:candidates} also shows the SNR per pixel and full width at half maximum (FWHM) (Columns 17 and 18).

\begin{figure*}
	\centering
	\includegraphics[width=1.0\textwidth]{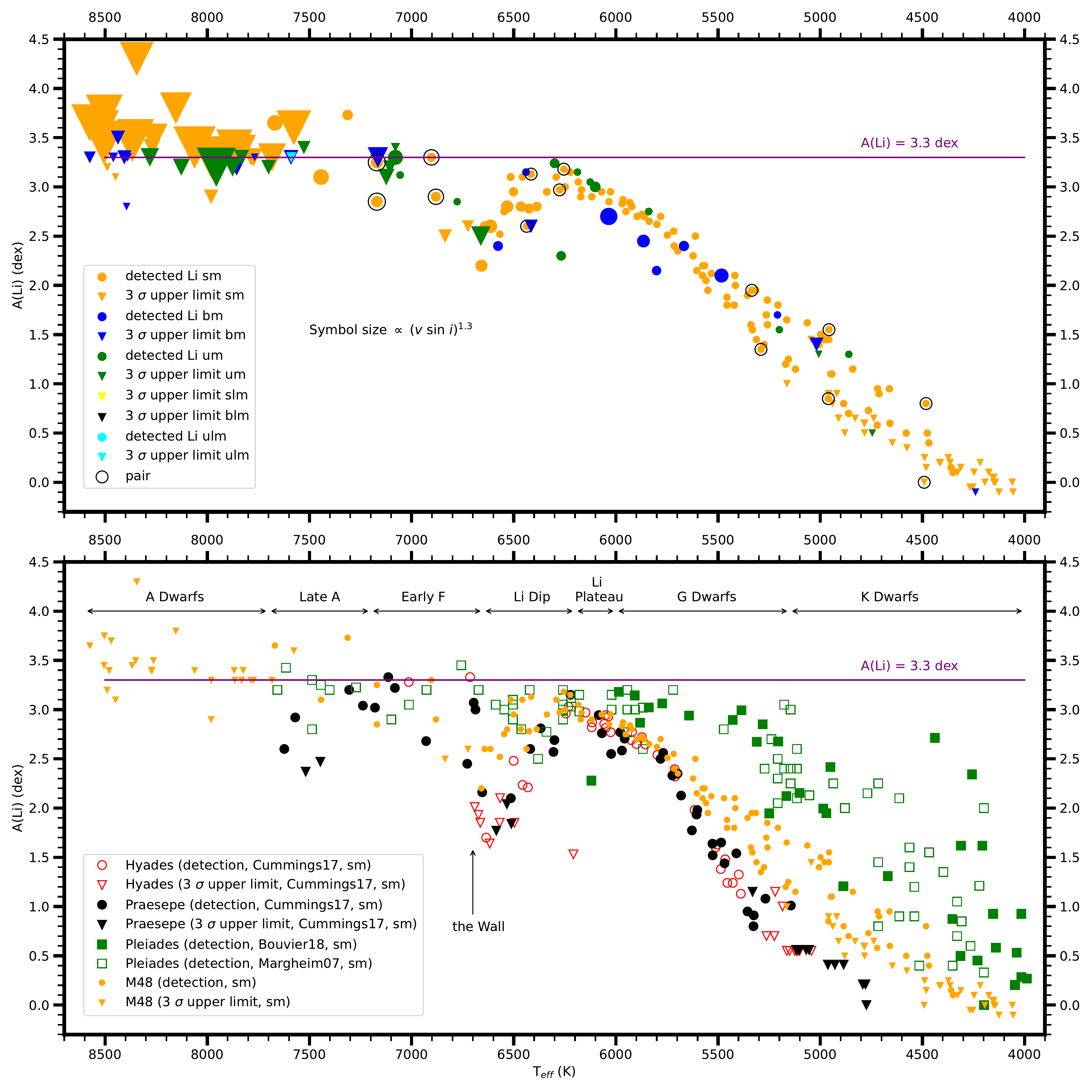}
	\caption{The A(Li) - $T_{\rm eff}$ patterns in M48. The top panel shows A(Li) for all members and likely members of M48, where different multiplicity/membership categories are shown using different colors.  The symbol sizes are proportional to ($v$ sin $i)^{1.3}$. The bottom panel shows M48 single members, only, compared to the sm from the Pleiades (120 Myr, from \citealt{2007PhDT.........2M} with supplemental data from \citealt{2018AA...613A..63B} for $T_{\rm eff}$ $<$ 6200 K) and the prime sample from the Hyades and Praesepe (C17).}
	\label{fig:ALi_teff}
\end{figure*}

Figure \ref{fig:ALi_teff} shows A(Li) versus $T_{\rm eff}$ for dwarf stars. In the top panel, all the cluster members and likely members (m and lm) are shown using different colors to indicate multiplicity/membership status. Detections are filled circles, upper limits are downward triangles, and the symbol sizes are proportional to ({\it v} sin {\it i})$^{1.3}$. Seven pairs of sm stars that have similar $T_{\rm eff}$ but different A(Li) (in each pair) are indicated by black open super-circles, and discussed below. As was found previously for the Hyades and Praesepe and discussed in \citet{1993ApJ...415..150T} and C17, binaries and stars with uncertain multiplicity and/or membership exhibit more scatter than single members. For example, {\it all} bm (dark blue) fall below the tight sm (orange) Li-$T_{\rm eff}$ trend for $T_{\rm eff}$ = 6100 - 5700 K, and ulm (light blue) lie almost consistently above the sm Li-$T_{\rm eff}$ trend for $T_{\rm eff}$= 6400 -- 6100 K. Since these deviations may obscure the true Li-$T_{\rm eff}$ trend for single members, we henceforth consider {\it only} the set of 171 single member stars. 

The bottom panel compares the sm (only) from M48 (age = 420 Myr) to single members of the Pleiades (age = 120 Myr) cluster and the C17 prime sample members of the Hyades and Praesepe (age = 650 Myr) clusters. Various features of the Li-$T_{\rm eff}$ relation are also marked and discussed below.  Finally, the meteoritic A(Li) is usually assumed to provide the initial solar A(Li), in agreement also with extended and presumably undepleted Li-$T_{\rm eff}$ plateaus observed in solar-metallicity clusters (\citealt{2011PhDT.......192C}).  \citet{1989GeCoA..53..197A} list meteoritic A(Li) = 3.31 $\pm$ 0.04 dex, whereas \citet{2021SSRv..217...44L} lists meteoritic A(Li) = 3.27 $\pm$ 0.03 dex. A purple line shows A(Li) = 3.3 dex. 

\begin{figure*}
	\centering
	\includegraphics[width=0.8\textwidth]{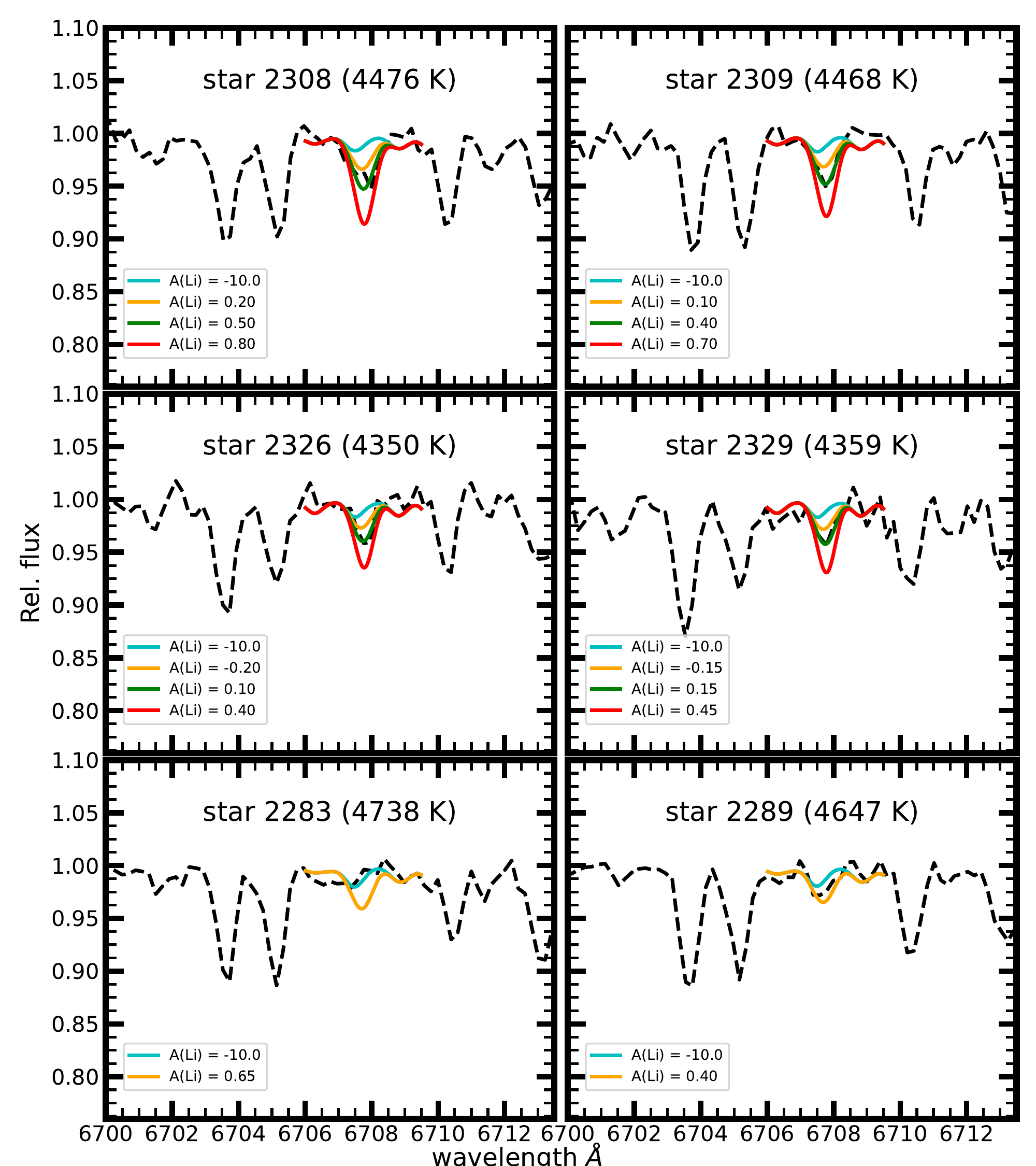}
	\caption{The observed and synthetic spectra for stars 2308, 2309, 2326, 2329, 2283, and 2289. In all panels, the cyan line shows synthesis for no Li (A(Li) = -10.0 dex). In the top and middle panels, the green line shows the best fit A(Li), and the orange and red lines show the A(Li) that are twice/half than the best fit A(Li). In the bottom panels, the orange line corresponds to the 3$\sigma$ upper limit A(Li).}
	\label{fig:cool_synth}
\end{figure*}

The numerous K dwarf Li detections help illuminate the rate of Li depletion during the MS and its possible connection to angular momentum loss (Section \ref{sec:discuss}).  Since detections can be more valuable than upper limits in this regard, we discuss a bit further the transition from mostly detections to mostly upper limits near 4500 K. Although the strength of the Li I line at constant abundance increases with lower $T_{\rm eff}$ , the combination of declining abundances and lower SNR for fainter stars means we lose the ability to detect Li in cooler K dwarfs.  Figure \ref{fig:cool_synth} shows the spectra of four of the five stars cooler than 4500 K that have Li detections; the fifth (star 2305) is shown in Figure \ref{fig:pair_comb}.  All five stars show clear absorption at the position of the Li I 6707.8 \AA\ line relative to the synthesis with no Li (blue line).  The best-fit synthetic A(Li) all provide good/excellent fits to the spectra and are all significantly higher than the 3$\sigma$ upper limit A(Li) calculated as indicated above. The errors in A(Li) (Column 20 of Table \ref{tab:candidates}) are those propagated from errors in both equivalent width and $T_{\rm eff}$. Since the portion due to equivalent width error is increasing in these fainter stars, the total errors themselves tend to be a bit larger than those for most other stars. Figure \ref{fig:cool_synth} and the bottom panel of Figure \ref{fig:star2161_2298} also show three stars with upper limits. Star 2283 clearly shows no evidence for absorption at the Li I position so we report the (conservative) 3$\sigma$ abundance as an upper limit. Stars 2289 and 2298 show hints of some absorption but since the possible absorptions are not significant at the 3$\sigma$ level we report 3$\sigma$ upper limits.

For some of the more subtle cluster comparisons it may be relevant to take into account the Galactic production of Li. Using extended Li-$T_{\rm eff}$ plateaus in young clusters of varying metallicity, \citet{2011PhDT.......192C} found evidence of Galactic Li production near solar metallicity with a production Li:Fe ratio close to 1:1. If this applies to the clusters considered here, then each cluster formed with a slightly different initial A(Li). So to study Li depletion we need to adjust all the stellar A(Li) to place them on a common scale for initial A(Li). Assuming solar metallicity as the reference, then we define

\begin{figure*}
	\centering
	\includegraphics[width=1.0\textwidth]{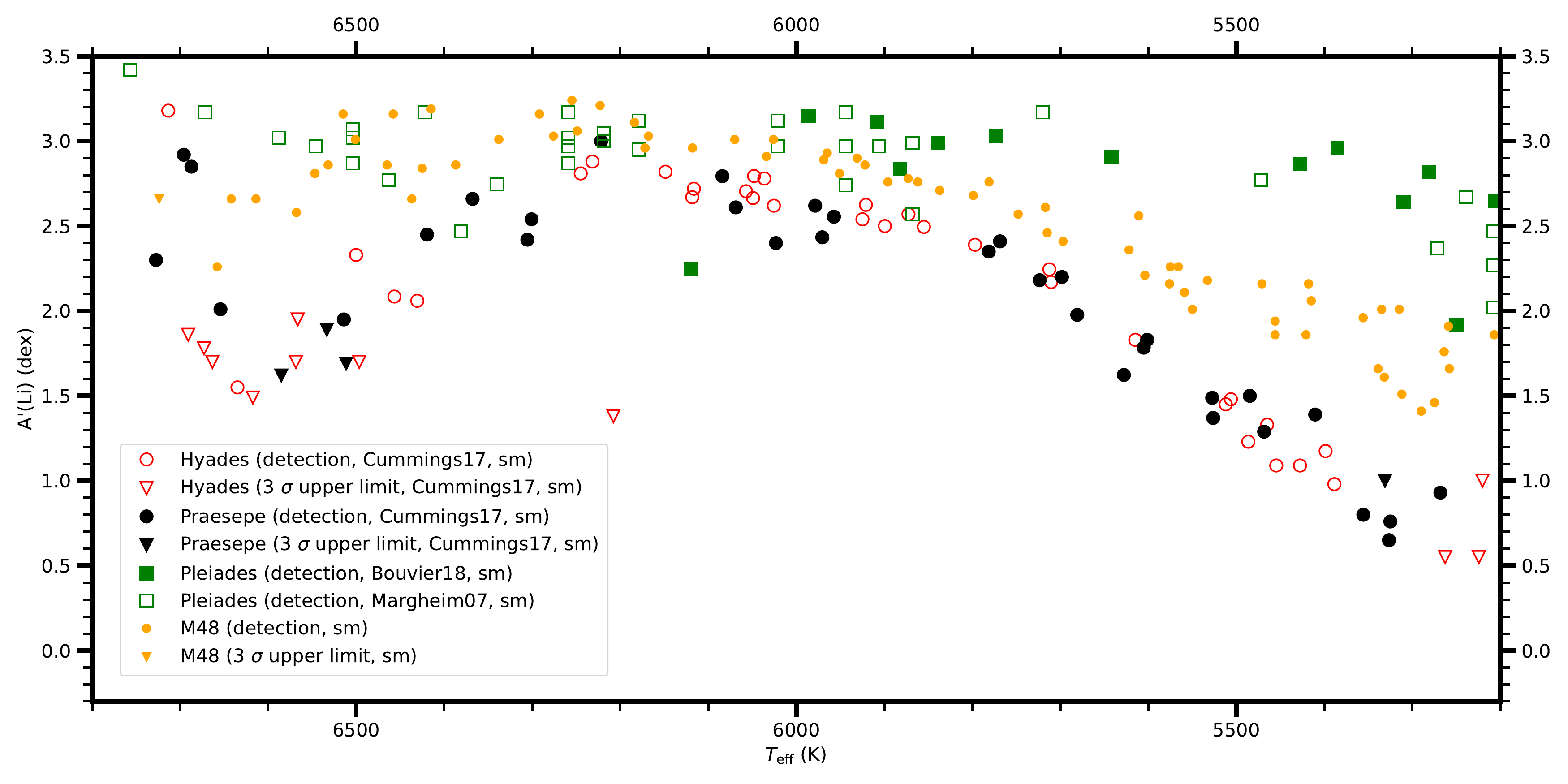}
	\caption{Magnification of a portion of the bottom panel of Figure \ref{fig:ALi_teff}. The A'(Li) scale is designed to allow for better comparison of Li {\it depletion} from cluster to cluster, by taking into account possible Galactic Li production effects on the initial cluster A(Li).  A production ratio of Li:Fe = 1:1 is assumed (\citealt{2011PhDT.......192C}). The reference A(Li) is assumed to be 3.30 at [Fe/H] = 0.0 dex.  Therefore, M48 data are shifted up from Figure \ref{fig:ALi_teff} by 0.06 dex, Pleiades data are shifted down by 0.03 dex, and Hyades/Praesepe data are shifted down by 0.15 dex.}
	\label{fig:ALi_shift}
\end{figure*}

A'(Li) = A(Li) - [Fe/H],

where A'(Li) is the shifted A(Li). For cluster [Fe/H] of $\sim$ -0.06 dex for M48 (Paper 1), $\sim$ +0.03 dex for the Pleiades (\citealt{2021ApJ...908..119M}), and $\sim$ +0.15 dex for the Hyades and Praesepe (C17), the cluster A(Li) are shifted by +0.06, -0.03, and -0.15 for M48, the Pleiades, and the Hyades/Praesepe clusters, respectively.  Figure \ref{fig:ALi_shift} zooms in on a portion of the bottom panel of Figure \ref{fig:ALi_teff}, where taking these shifts into account may be important.

\section{discussion} \label{sec:discuss}

In this section, we discuss the A(Li) -- $T_{\rm eff}$ pattern of M48 dwarfs (and 3 post-MS stars), compare to the Pleiades and Hyades/Praesepe clusters, and discuss possible interpretations.

\subsection{Comparison of Li-$T_{\rm eff}$ Trends in M48, Pleiades, Hyades, and Praesepe} \label{sec:trend}

In studying how the dwarf Li-$T_{\rm eff}$ trend forms and evolves to the age of M48 and beyond, it is convenient to start with the relatively young Pleiades cluster. Overall, the Li-$T_{\rm eff}$ trend seems relatively simple, going from near-meteoritic A(Li) in hotter stars to increasingly depleted A(Li) in cooler stars.  However, some complexities already appear: some A, F, and early G stars show scatter (for example near 7500, 6300, 6100, and 5800 K), and increasingly large scatter accompanies the dramatically increasing Li depletion in late G and K dwarfs.

In the Hyades and Praesepe clusters, the dwarf Li-$T_{\rm eff}$ relation is strikingly different and more complex than that in the Pleiades. C17 found that the Hyades and Praesepe clusters are indistinguishable in age, composition, and Li-$T_{\rm eff}$ relation, and their Li-$T_{\rm eff}$ relations complement each other, so it is useful to use the combined Hyades/Praesepe Li-$T_{\rm eff}$ relation as a single relation. Following the Li-$T_{\rm eff}$ trend once more from higher to lower mass, late A stars are clearly depleted relative to the Pleiades. There is a severe depletion of Li for stars with 6675 K $<$ $T_{\rm eff}$ $<$ 6200 K (the ``Li Dip", first discovered in the Hyades by \citealt{1986ApJ...302L..49B}) which is absent or near-absent from the Pleiades (\citealt{1988ApJ...327..389B}). The C17 combined Hyades/Praesepe sample refined knowledge of the Li Dip: from 6200 K, increasing $T_{\rm eff}$ implies steadily increasing Li depletion with little scatter, until 6635 K where 9 stars in a {\it tiny} range of $T_{\rm eff}$ (of $<$ 100 K) show A(Li) increasing by more than 1.6 dex (the ``Wall").  Slightly cooler stars show a Li Plateau that is arguably lower than that in the Pleiades, and G and K dwarfs show far greater Li depletions than in the Pleiades.

These patterns suggest classification of the different features of the Li-$T_{\rm eff}$ relation as follows (see also C17 and \citealt{2000ASPC..198..235D}): A dwarfs (8600 - 7700 K), late A dwarfs (7700 - 7200 K), early F dwarfs (7200 - 6650 K), the Li Dip (6675 - 6200 K), the Li Plateau (6200 - 6000 K), G dwarfs (6000 - 5150 K), and K dwarfs (5150 - 4000 K).  Note the Wall at $T_{\rm eff}$= 6700 K.

Even a quick glance at Figure \ref{fig:ALi_teff} suggests the presence of all these features and perhaps a few more in M48. Perhaps not surprisingly, the above features appear to be at a stage of development that is intermediate to that in the Pleiades and the Hyades/Praesepe clusters. We now delve into each feature in more detail.

\subsubsection{A dwarfs ($>$7700 K)}

Detecting Li in stars that are expected to be rotating rapidly and where the Li line is expected to be very weak can be quite challenging. We have nevertheless observed as many stars as was feasible in case there were discoverable surprises. Unfortunately, nearly all stars have upper limits near or above the meteoritic A(Li) (Figure \ref{fig:ALi_teff}), so it is not possible to discern whether these stars are depleted in Li or even slightly enriched. Although star 3010 has A(Li) $<$ 2.9 dex, we caution against concluding that it is depleted.  Note that the Hyades and Praesepe Am stars of \citet{1989A&A...220..197B, 1998A&A...338.1073B} were not included in the prime sample of C17.

\subsubsection{Late A (7700 -- 7200 K)}

Several Pleiads show a plateau near 3.3 dex, consistent with the meteoritic A(Li) and an assumption of being undepleted. Most Praesepe stars and possibly one Pleiad are depleted relative to the Pleiades trend and to M48. Of the five M48 stars in this $T_{\rm eff}$ range, at least two stars (2080, 2091) are extraordinary in exhibiting significant Li enrichment above 3.3 dex at levels 3.65 - 3.75 dex. We cannot discern whether star 2083 (A(Li) $<$ 3.6 dex) is also extraordinary. It may be interesting that whatever physical processes that have enriched Li in the M48 stars have not affected the Pleiades or Praesepe.

\subsubsection{Early F (7200 -- 6735 K, excluding the Wall)}

\begin{figure}
	\centering
	\includegraphics[width=0.45\textwidth]{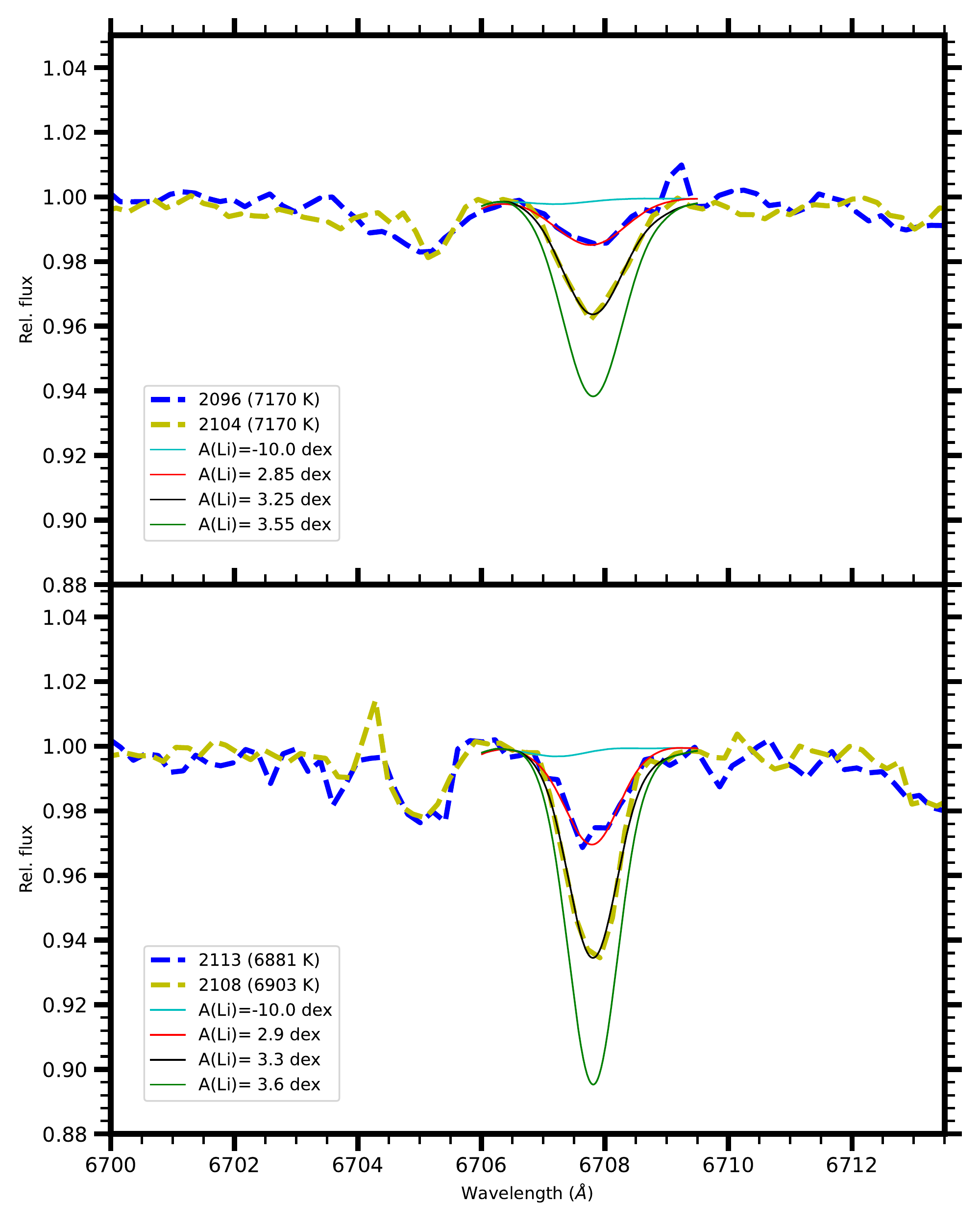}
	\caption{Comparisons between A(Li) for pairs of stars with similar $T_{\rm eff}$. In each panel, the blue and yellow dotted lines are the observed spectra for each pair. Also shown are the condition of no Li (cyan line), best fit A(Li) for the two stars (red and black lines), and A(Li) that is two times larger than the black line.}
	\label{fig:hot_Lidip}
\end{figure}

In the region that is just slightly hotter than the Li Dip (6850 - 7200 K), three Hyades/Praesepe stars have detections near 3.3 dex, and two are slightly depleted ($\sim$ 3.0 and 2.7 dex).  Interestingly, one or maybe two Pleiads may also be slightly depleted. Four of the five M48 stars comprise two interesting pairs where each pair has nearly identical $T_{\rm eff}$ but different A(Li). Figure \ref{fig:hot_Lidip} shows spectra of and syntheses for the two pairs. In each pair the Fe I lines near 6705 line up well, but the Li lines are clearly different. In both pairs, the higher A(Li) is consistent with meteoritic but the other star is lower by 0.4 dex, illustrating clear Li depletion. The upper limit of star 2119 (A(Li $<$ 2.5 dex) also suggests depletion. Possibly so does star 2113, A(Li) $<$ 2.9 dex.  While Li depletion in this mass range has been observed in substantially older clusters (for example in NGC 7789, NGC 3680, and NGC 6819 with ages 1.5 - 2.2 Gyr, \citealt{2019AJ....158..163D}), these examples show that Li depletion begins at least as early as 420 Myr.

\subsubsection{Li Dip and the Wall (6735 -- 6200 K)}

Figure \ref{fig:ALi_teff} clearly shows the formation and evolution of the Li Dip in going from the Pleiades to M48 to the Hyades/Praesepe. In the Pleiades, a few stars may already be showing significant depletion (A(Li) = 2.5 - 2.9 dex) but most stars show at most a small depletion (if any) relative to the two stars near 6700 K that straddle the meteoritic A(Li). 

The Hyades/Praesepe prime sample of C17 define a very precise Li Dip. In the $T_{\rm eff}$ range 6200 - 6635 K (the ``cool" or ``red" side of the Li Dip), there is increasing Li depletion with little scatter. Then, remarkably, in a range of less than 100 K (from 6635 to 6735 K) there is an apparent very steep rise (``the Wall", or the ``hot" or ``blue" side of the Li Dip) defined by six Hyades/Praesepe detections from A(Li) = 1.70 to 3.33 dex, and enhanced by three upper limits between 1.85 and 2.01 dex. Note that the two Pleiads mentioned above lie at the top of the Wall. A similar vertical structure is seen the older clusters NGC 752 (1.45 Gyr, \citealt{2022ApJ...927..118B}), NGC 3680 (1.75 Gyr, \citealt{2009AJ....138.1171A}), and the much richer samples of NGC 7789 (1.5 Gyr, \citealt{2020MmSAI..91...74T}) and NGC 6819 (2.2 Gyr, \citealt{2019AJ....158..163D}) that separates hotter stars that show a large variety of A(Li) (from 3.3 to $<$ 1.6 dex) from Li Dip stars whose A(Li) are uniformly $<$ 2.4 dex across the hotter 400 K in portion of the Li Dip. Unfortunately only four M48 stars fall in this $T_{\rm eff}$ range so the situation in M48 is unclear. Two are detections at A(Li) = 2.2 and 2.6 dex while the other two are upper limits at 2.5 and 2.6 dex. 

\begin{figure}
	\centering
	\includegraphics[width=0.45\textwidth]{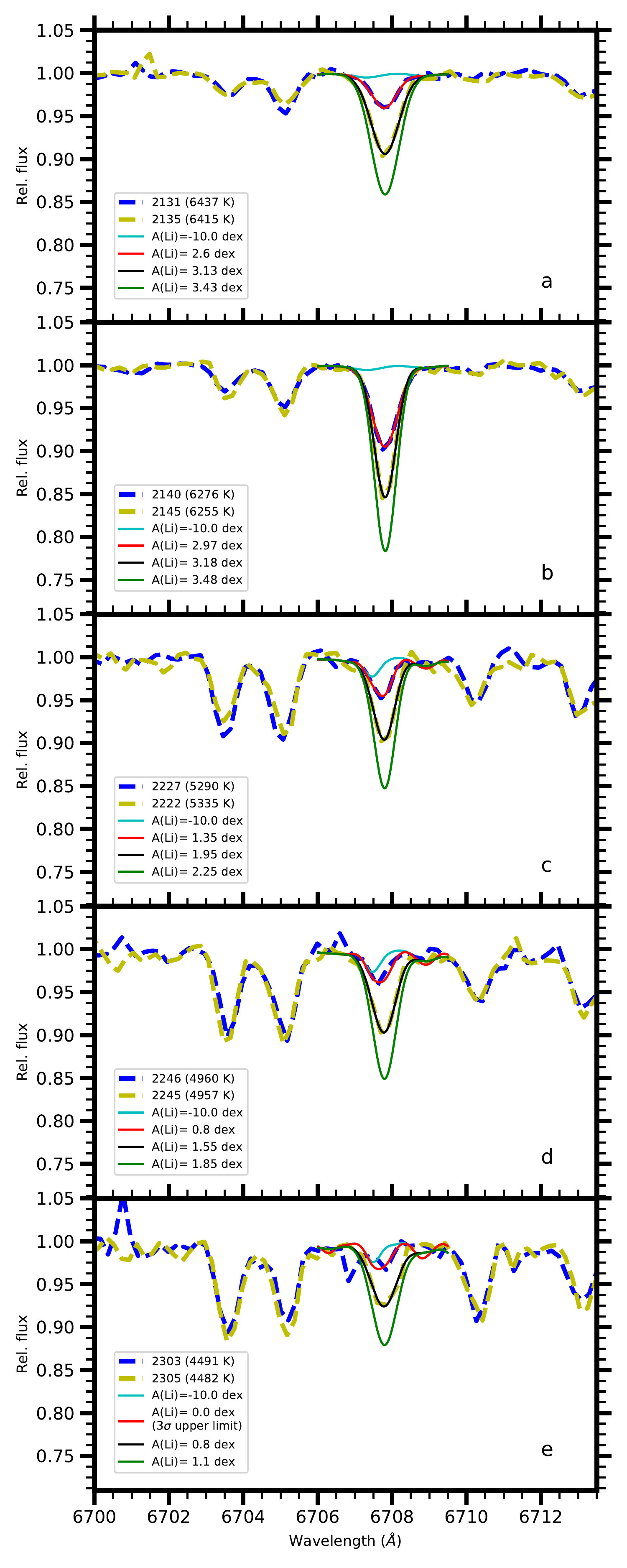}
	\caption{Comparisons between A(Li) for pairs of stars with similar $T_{\rm eff}$, with $T_{\rm eff}$ decreasing from top to bottom. In each panel, the blue and yellow dotted lines are the observed spectra for each pair. Also shown are the condition of no Li (cyan line), best fit A(Li) for the two stars (red and black lines), and A(Li) that is two times larger than the black line.}
	\label{fig:pair_comb}
\end{figure}

All M48 stars on the cool side of the Li Dip have detections and generally lie between the Pleiades and the Hyades/Praesepe. Like the Hyades/Praesepe, the cool side of the M48 Li Dip shows a clear pattern of increasing Li depletion with increasing $T_{\rm eff}$, although with more scatter than in the Hyades/Praesepe. Figure \ref{fig:pair_comb}a shows a pair near 6400 K with A(Li) that differ by 0.55 dex. Either such a difference becomes smaller by the age of the Hyades/Praesepe, or the Hyades/Praesepe is unusual, or the samples are too small to fully describe how Li Dip stars act.

\subsubsection{Li Plateau (6200 - 6000 K)}

The Li plateau of the Pleiades shows very little slope, if any, but may be slightly depleted, as compared to the hotter stars discussed above and to the meteoritic A(Li).  The increasing slope from the hottest dwarfs through to G (and K) dwarfs is consistent with SSET, as discussed earlier.  However, the older clusters show a slope in the Plateau, which steepens further for $T_{\rm eff} <$ 6000 K, so we limit our definition of the Li Plateau from about 6200 to 6000 K. SPTLBs provide strong evidence that the Li Plateaus in the Hyades and in M67 are depleted (Section \ref{sec:intro}) but any differences in the Plateaus of the clusters discussed here are subtle. M48 and the Hyades/Praesepe appear to be lower than the Pleiades, but not necessarily distinct from each other. However, if Galactic Li production affected the initial abundances of these clusters (end of Section \ref{sec:li}), then to study the relative Li depletion of these clusters we must place them on a common scale for initial A(Li). In particular, relative to the Sun, the A(Li) of all Pleiads must be lowered by 0.03 dex, all Hyads and Praesepids must be lowered by 0.15 dex, and all M48 members must be increased by 0.06 dex. Figure \ref{fig:ALi_shift} shows a close-up of a portion of the bottom panel of Figure \ref{fig:ALi_teff} using this adjusted scale for A(Li), for the regions where this issue is most relevant. Under these assumptions, the Hyades/Praesepe Li Plateau is clearly lower than the one in M48, and the Pleiades Li Plateau remains slightly above M48. So here too, we see a progression of Li depletion from the Pleiades to M48 to the Hyades/Praesepe.

Just hotter than our defined boundary for the Li Plateau (near 6257 K) there is evidence for some scatter, as illustrated in Figure \ref{fig:pair_comb}b.  (Figures \ref{fig:ALi_teff} and \ref{fig:ALi_shift} show that the Pleiades might also show scatter at this same $T_{\rm eff}$.)  However, the M48 Li Plateau itself is very tight with remarkably little scatter.  

\subsubsection{G/K dwarfs (6000 - 5150 K, 5150 - 4000 K)} \label{sec:GK dwarf}

The Li-$T_{\rm eff}$ trend in M48 lies significantly below that of the Pleiades and slightly above that of the Hyades/Praesepe. We argue below that the dominant factor is increasing depletion with age; however, metallicity may also play a role. In addition, whereas the early G stars show remarkably tight Li-$T_{\rm eff}$  relations, cooler stars show large scatter, as exemplified by direct comparisons of spectra of 3 pairs of stars in Figure \ref{fig:pair_comb}c,d,e.  Each pair has nearly identical $T_{\rm eff}$ (near 5300, 4960, and 4485 K) but significantly different A(Li) (0.6, 0.75, and $>$ 0.8 dex, respectively). 

Standard Li depletion of G/K dwarfs increases with increasing metallicity (e.g. P97).  Since M48 is {\it metal-poor} compared to the Pleiades, age may be the dominant parameter affecting the difference in the Li-$T_{\rm eff}$ trends of these clusters.

In the Pleiades (\citealt{1987ApJ...319L..19B, 1993AJ....106.1059S, 2018AA...613A..63B}) and in the slightly older M35 (\citealt{2018AJ....156...37A, 2021MNRAS.500.1158J}), ultra-fast rotators (UFRs) exhibit significantly higher A(Li) than slower rotators, at least in part creating the large Li dispersions seen in these clusters. If M48 formed with its share of UFRs, they would appear to have all spun down (\citealt{2015AA...583A..73B}); all of our G and K dwarfs have {\it v} sin {\it i} $< 20$ km s$^{-1}$. It would be nice to know what happens to the A(Li) of UFRs as they spin down, but clusters with ages closer to those of the Pleiades and M35 may be required to address this. M48 itself provides at best a mild hint: We have detected Li in {\it all} M48 stars in the $T_{\rm eff}$ range 5400 - 5050 K, which is the upper part of the $T_{\rm eff}$ range where high-Li UFRs appear in the Pleiades and M35, and these M48 stars show a Li dispersion of no more than 0.6 - 0.7 dex. By contrast, the dispersions in the Pleiades and M35 near 5300 - 5200 K are at least 1.3 dex, so it is possible that the Li dispersions in this $T_{\rm eff}$ range has shrunk by the age of M48. Cooler stars in the Pleiades and M35 show even larger Li dispersions, at least 1.8 dex in both clusters, but because our M48 data include upper limits for $T_{\rm eff}$ $<$ 5050 K, we cannot evaluate the full dispersion range for cooler M48 stars. All we can say is that it is at least 0.7-0.8 dex near 5000 K and 4500 K.

Comparison of Li depletion in M48 to that in the Hyades/Praesepe is complicated by the difference in [Fe/H] of $\sim$ 0.21 dex. As discussed above, Galactic Li production suggests the Hyades/Praesepe A(Li) should be moved down relative to M48 by about 0.21 dex, in which case the entire G/K dwarf Li-$T_{\rm eff}$ trend in the Hyades/Praesepe lies clearly below that of M48 (Figure \ref{fig:ALi_shift}). But which is more important, age or metallicity (or something else)? By way of example, SSET models of P97 show a $\sim$ 0.4 dex difference in Li depletion between [Fe/H] of -0.06 dex (M48) and +0.15 dex (Hyades/Praesepe) at 5500 K. The Galactic-production-adjusted difference between the Li-$T_{\rm eff}$ trends of these clusters at 5500K is 0.66 dex, suggesting that the difference between the two is partly due to metallicity and partly to age. However, \citet{2014ApJ...790...72S} highlight a number of uncertainties that affect standard depletion, some of which (like opacity) are themselves metallicity-dependent. So the relative effects of metallicity and age on Li depletion could be a bit different than indicated here. But what does seem clear is that age-related Li depletion is far greater between the Pleiades and M48 than it is between M48 and the Hyades/Praesepe. Since the age differences are similar, where M48 is about 300 Myr older than the Pleiades and about 250 Myr younger than the Hyades/Praesepe, main sequence G/K dwarf Li depletion slows down over time.

\subsection{Interpretations}

In this section we discuss possible interpretations of the various observed features enumerated in Section \ref{sec:trend}. The discussion proceeds from higher to lower mass, so we begin with the giants.

\subsubsection{Giants}  \label{sec:giants}

SSET predicts that as stars evolve past the turnoff to lower $T_{\rm eff}$ and become subgiants, the SCZ deepens past the boundary of the Li preservation region, thereby diluting the surface A(Li) (\citealt{1967ApJ...147..624I, 1990ApJS...73...21D}). The SCZ reaches a maximum depth by mass fraction, resulting in maximum dilution, after which it retreats, leaving behind a diluted Li-$T_{\rm eff}$ plateau. Standard dilution alone can explain the difference between the turnoff A(Li) and the diluted plateau in metal-poor stars (e.g. \citealt{1995ApJ...453..819R}). However, open clusters show a more complex pattern (\citealt{2009AJ....138.1171A, 2018AJ....155..138A, 2021AJ....161..159A, 2012AJ....144..137C, 2015ApJ...799..202B, 2019AJ....158..163D}). Near solar metallicity, decreasing giant mass (older cluster age) shows increasing departure from SSET, perhaps largely due to Li depletion during earlier phases. Decreasing metallicity suggests closer conformity to SSET.

\begin{figure}
	\centering
	\includegraphics[width=0.45\textwidth]{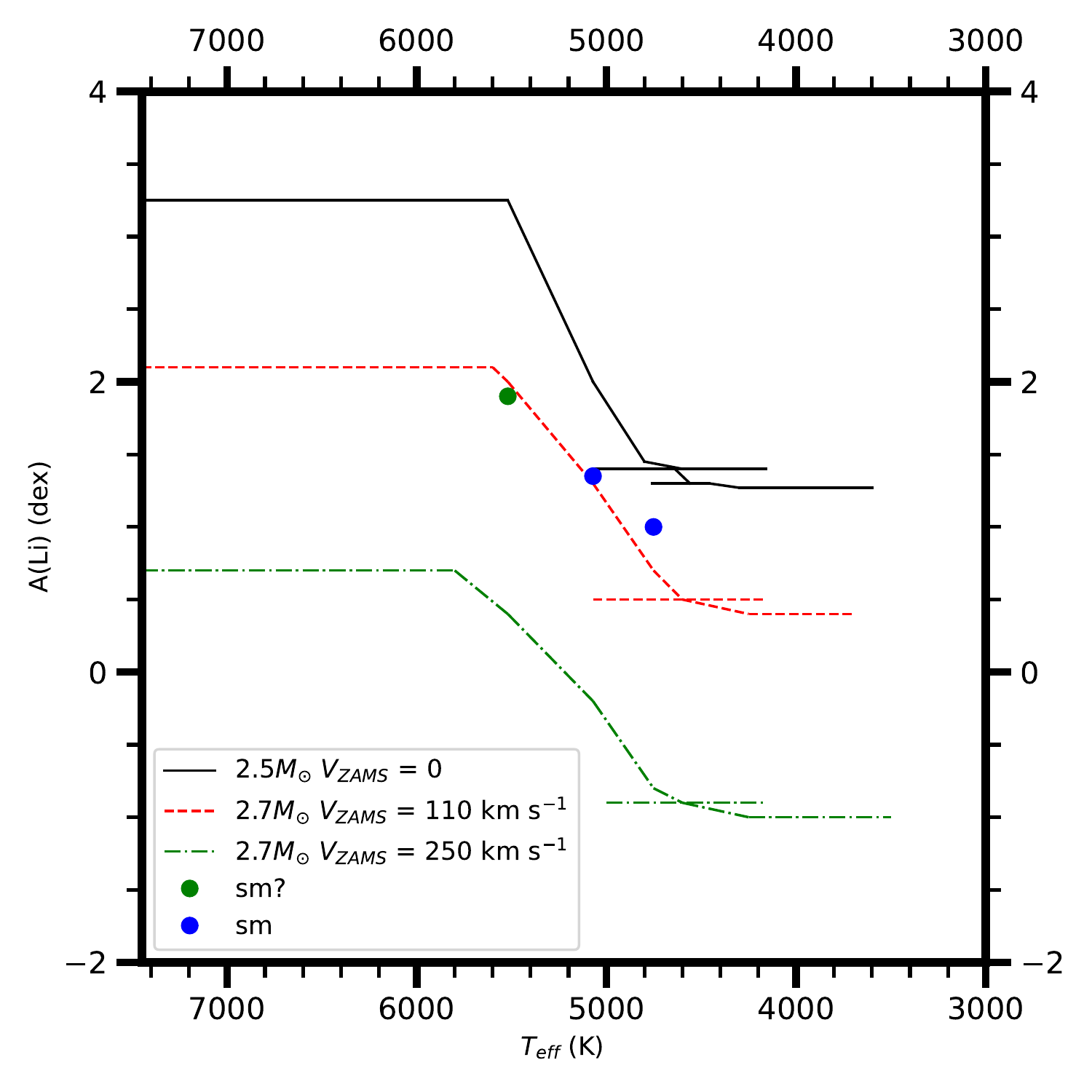}
	\caption{Post-MS Li evolution tracks for 2.5 - 2.7 $M_{\odot}$ giants with $V_{ZAMS}$ of 0 (SSET), 110, and 250 km s$^{-1}$. Star 2002 is marked with green disk and stars 2003-4 are marked with blue disks.}
	\label{fig:giant}
\end{figure}

Since M48 has very few giants, we make an exception and consider the ``slm" giant (star 2002) in addition to the two ``sm" giants (stars 2003-4). Using a 420 Myr, [Fe/H] = -0.063 dex $Y^2$ isochrone (\citealt{2001ApJS..136..417Y}), we estimate our giants masses to be $\sim$ 2.9 $M_{\odot}$. Of particular relevance is the right panel of Figure 14 from \citet{2010A&A...522A..10C}, which shows post-MS Li evolution tracks for 2.5 - 2.7 $M_{\odot}$ giants with ZAMS equatorial velocities ($V_{ZAMS}$) of 0 (SSET), 110, and 250 km s$^{-1}$ (these models are also shown in our Figure \ref{fig:giant}). The standard track leaves the MS with no Li depletion, whereas the other two leave with depletions of 1.15 and 2.55 dex, respectively, due to rotational mixing. All three begin dilution near 5600 - 5500 K (a similar $T_{\rm eff}$ as the metal-poor models of \citealt{1995ApJ...453..819R}), and dilute by 1.85, 1.55, and 1.55 dex, respectively (factors of 71, 35, and 35) down to 4800 - 4600 K.

In Figure \ref{fig:giant}, star 2002 is shown in filled green disk (slm), to be distinguished from 2003-4 in filled blue disks (sm). Star 2002 is remarkable in that its $T_{\rm eff}$ of 5520 K places it in the middle of the Hertzsprung gap, right at the boundary of where model dilution begins. The depleted A(Li) of 1.9 dex is consistent with MS Li depletion of the model with $V_{ZAMS}$ = 110 km s$^{-1}$, or just slightly larger.  Star 2004 ($T_{\rm eff}$ = 5071 K) is consistent with being in the middle of dilution, needing also MS depletion from the same 110 km s$^{-1}$ track as 2002. Star 2003 ($T_{\rm eff}$ = 4752 K) is consistent with being near the end of dilution and would require a slower rotator than the previous two, perhaps near 50 km s$^{-1}$. It is possible that stars 2004 and 2003 are more evolved red clump stars that have suffered an additional 0.1 - 0.2 dex depletion between the RGB and the red clump, according to the models of \citet{2010A&A...522A..10C}. In that case, a model with $V_{ZAMS}$ = 0 - 20 km s$^{-1}$ could match star 2004 whereas 30 - 40 km s$^{-1}$ could match star 2003. Except for this one possibility with star 2004, the other scenarios discussed above require approximately 0.5 - 1.0 dex Li depletion during the MS. Note that additional internal effects such as thermohaline mixing become important only for significantly lower model masses.

It is possible to envision more complex possibilities (see Section \ref{sec:lateA}). For example, faster rotators may have depleted more Li, and then enrichment maybe have occurred through accretion/engulfment. Such a scenario may be discernible through abundance signatures other than Li.

\subsubsection{Late A}  \label{sec:lateA}

Possible explanations for the two Li-rich stars with A(Li) = 3.65 and 3.73 dex include diffusion (radiative levitation), planetesimal accretion, and engulfment of a planet or Li-preserving brown dwarf. Each scenario has advantages and disadvantages. 

Diffusion is more efficient in slower rotators, and our two stars indeed have {\it v} sin {\it i} of 33 and 60 km s$^{-1}$, which is slower than most of our hotter stars and a few stars slightly cooler. Two other stars in the same $T_{\rm eff}$ range rotate much faster and have upper limits in A(Li) of 3.6 dex, which is uninformative, and 3.3 dex, which is more convincingly below the two Li-rich stars. Star 2095 has a detection of A(Li) at 3.1 dex, which is near though perhaps marginally below meteoritic, and a {\it v} sin {\it i} of 61km s$^{-1}$. However, Li-richness in the models of \citet{1993ApJ...416..312R} at this age is constrained to the narrow $T_{\rm eff}$ range of 6950 - 7100 K, the ``Li Peak", which is clearly cooler than our stars. On the other hand, the super-Li-rich dwarf J37 of NGC 6633 may lie in this $T_{\rm eff}$ range (\citealt{2002ApJ...577L..39D}). It was suggested that J37 might be better explained by planetesimal accretion (\citealt{2003ApJ...595.1148L}), but the overall abundance pattern of a number of elements fit neither scenario conclusively (\citealt{2005MNRAS.363L..81A}).

A dwarfs have extremely thin SCZs so it only takes accretion of just a small amount of hydrogen-poor material to enhance the surface abundances of the accreted elements. Accretion of planetesimals and planet engulfment are two possibilities for such enhancement. However, given how easy it is to produce such enhancements, one wonders whether Li-rich A dwarfs are observed sufficiently often. In M48, two out of five stars in this $T_{\rm eff}$ range are Li-rich, but we would expect to see Li-rich stars from accretion in hotter stars as well, yet most of our hotter M48 stars have A(Li) $<$ 3.5 dex. Another issue is that rotational mixing may deplete Li, as suggested for the M48 giants (Section \ref{sec:giants}) and for stars a bit more massive than the Li Dip (\citealt{2019AJ....158..163D}).  So accretion may enrich the surface Li and subsequent depletion may diminish it. Finally, some A dwarfs show a number of abundance anomalies, at least some of which might be explained by diffusion. Fortunately, the signatures of diffusion and accretion differ, so it may be possible to distinguish between these mechanisms by studying the abundances of appropriate elements; for example, accretion may show a pattern of more refractory elements than the volatiles (\citealt{2020ApJ...888L...9N}).

\subsubsection{Li Dip}

As stars evolve, differential contraction together with loss of angular momentum from the surface can trigger a secular shear instability (and others) that causes mixing and depletion of the surface Li, Be, and B (\citealt{1990ApJS...74..501P, 1995ApJ...441..876C, 1997ApJ...488..836D, 2016ApJ...830...49B}). Much evidence from a variety of angles favors such rotational mixing as the dominant Li-depletion mechanism in the Li Dip, as discussed in Section \ref{sec:intro}.  In addition, since stars spin down (e.g. \citealt{2007ApJ...669.1167B}) as the Li Dip develops (e.g. from the Pleiades to M48 to the Hyades/Praesepe), there is a correlation between stellar spindown and Li depletion. Furthermore, the existence of Li dispersions at a given $T_{\rm eff}$ separates proposed mechanisms that are able to create such dispersions, such as the above rotational mixing, from those that cannot, such as diffusion and gravity waves. Our M48 data help define the evolution of the Li Dip and constrain the degree of Li dispersion in the Li Dip, which can help guide future models.

In principle, diffusion can occur in sufficiently stable stellar layers. It is thus possible that diffusion contributes to Li depletion in the very small $T_{\rm eff}$ range (of about 100 K near 6700 K, depending also slightly on age, \citealt{1993ApJ...416..312R}) where it is predicted to be significant. Diffusion is predicted to act differently on Be than rotational mixing. While both elements are fully ionized at the base of the SCZ ($T_{\rm eff}$ $<$ 6650 K) they diffuse downwards via gravitational settling at similar rates; the patterns become more complex at higher $T_{\rm eff}$ as one or both elements are able to retain at least one electron at the base of the SCZ (e.g. Figure 12 of \citealt{1993ApJ...416..312R}). The preponderance of Li and Be data in the Li Dip (\citealt{1997ApJ...491..339S, 1998ApJ...498L.147D, 2001ApJ...553..754B, 2004ApJ...613.1202B}) are more consistent with the signature of rotational mixing. However, HR 6052 is an interesting exception (\citealt{1997ApJ...491..339S}): at $T_{\rm eff}$ = 6712 K, it has A(Li) = 2.82 dex and A(Be) = 0.48 dex, which are 0.49 and 0.94 dex below meteoritic, respectively. Such a depletion pattern is inconsistent with rotational mixing, diffusion, or accretion, or combinations there of. The diffusion models of \citet{1993ApJ...416..312R} have significant Be depletion at a $T_{\rm eff}$ slightly higher than 6850 K, but the A(Li) is then higher than meteoritic due to radiative levitation. Interestingly, another star may be closer to this diffusion pattern: at $T_{\rm eff}$  = 6953 K, HR 1287 has A(Li) = 3.42 dex (near or slightly super-meteoritic) and A(Be) = 0.89 dex (sub-meteoritic, \citealt{2001ApJ...553..754B}). It would be worthwhile to greatly expand the sample of stars in the $T_{\rm eff}$ range 7000 - 6600 K that have measurements of both A(Li) and A(Be).

\subsubsection{Li Plateau; G/K dwarfs} \label{sec:GK}

Higher A(Li) in SPTLBs as compared to normal, single stars provide direct support for models with rotational mixing related to angular momentum loss. Here too, stellar spindown correlates with Li depletion, and Li dispersions distinguish between models that can create such dispersions (rotational mixing) from those that cannot (SSET, including with convective overshoot; diffusion; and gravity waves). Convective overshoot for ZAMS-defined G dwarfs occurs during the pre-MS only, in contradiction to the degree of Li depletion seen in the Pleiades and to the MS Li depletion inferred from other clusters (including M48), while mass loss leads to various absurdities (see discussion in C17). Some diffusion may be possible, though helioseismology limits it to less than 0.1 dex in the Sun and it would be even smaller for cooler dwarfs. It is possible that mixing by gravity waves may play a role in addition to rotational mixing; however the timing of the Li depletion places constraints. Li depletion slows down over time (Section \ref{sec:GK dwarf}) as does stellar spindown, but mixing by gravity waves is more closely related to action of the base of the SCZ, whose depth remains more steady with age. Perhaps mixing by gravity waves becomes more important at older ages.

Given the preponderance of evidence suggesting that rotationally-induced mixing through angular momentum loss (hereafter J-loss) is the dominant Li-, Be-, and B- depleting mechanism in dwarfs spanning a variety of spectral types, including G/K (C17, Section 1), we examine our data for possible connections between J-loss and Li depletion in G/K dwarfs.

\begin{figure*}
	\centering
	\includegraphics[width=0.95\textwidth]{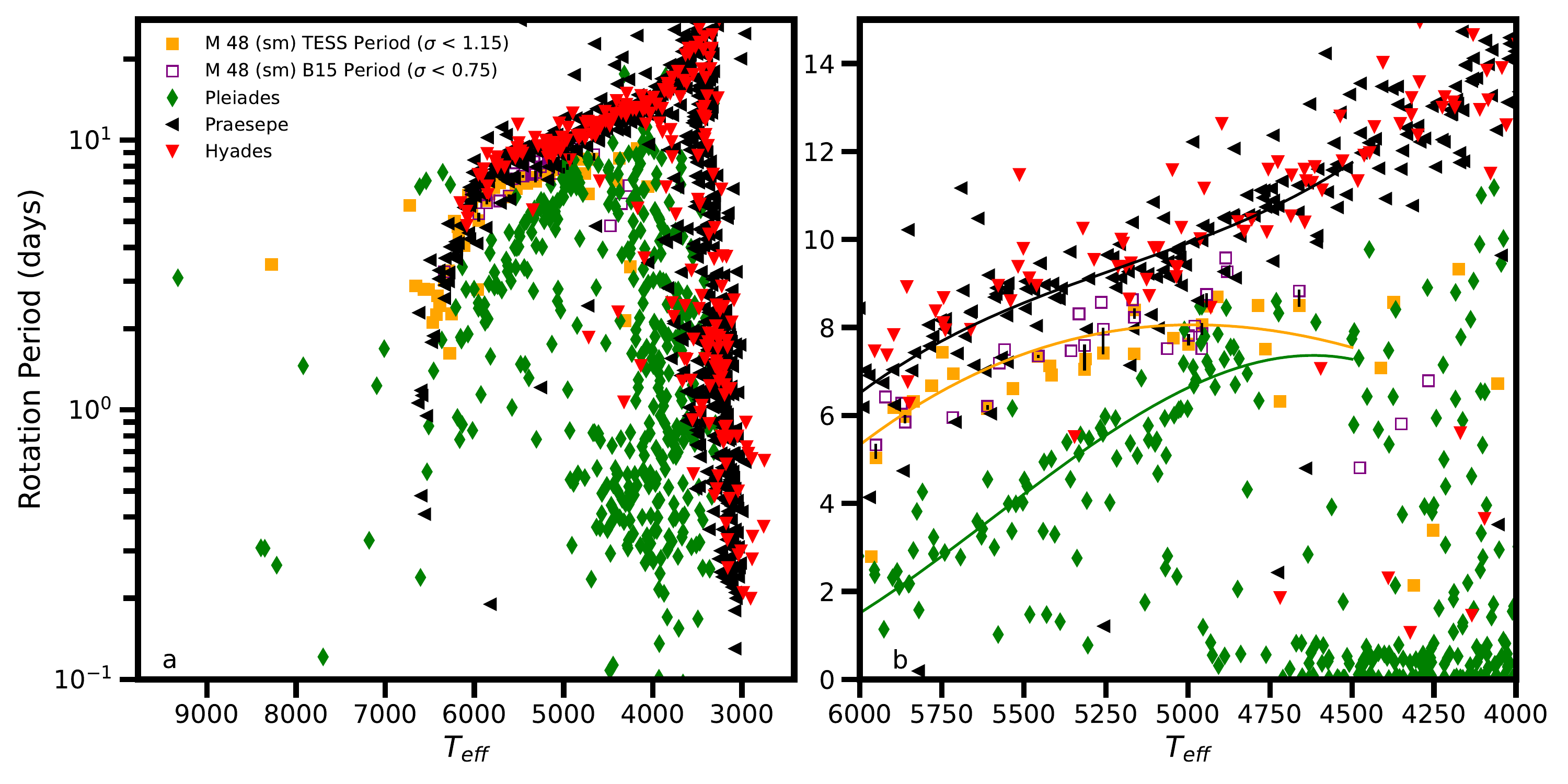}
	\caption{The rotational period -- $T_{\rm eff}$ patterns in M48 (only sm shown), Pleiades (green, from \citealt{2016AJ....152..113R}), Hyades (red, from \citealt{2019ApJ...879..100D}), and Praesepe (black, from \citealt{2021ApJ...921..167R}). The left panel includes the entire $T_{\rm eff}$  range with reported periods (y axis in log scale), and the right panel zooms in to the 6000 - 4000 K region (y in linear scale). For M48, the TESS periods (see text) are filled orange squares and the B15 periods are open purple squares; the short dashes connect stars that have both periods. For M48, only stars with well-measured periods are included, as discussed in Section \ref{sec:rad}. The curves are fits to the general trend from 6000 to 4500 K (black = Hyades + Praesepe; orange = M48; and green = Pleiades slower rotators).}
	\label{fig:LinearPeriods}
\end{figure*}

\begin{figure*}
	\centering
	\includegraphics[width=0.6\textwidth]{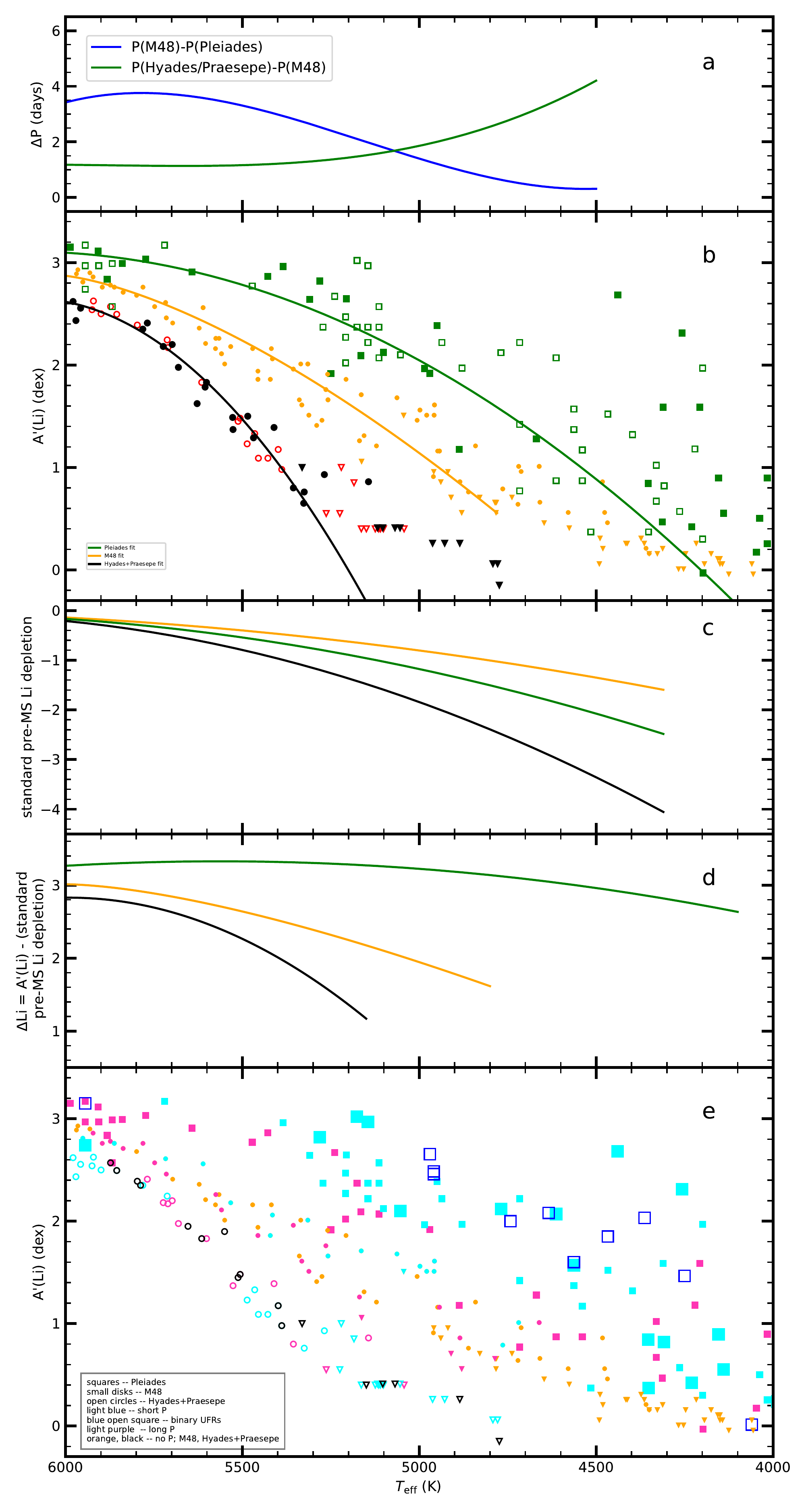}
	\caption{From top to bottom: a) Differences, $\Delta$P, in the cluster fits to the periods from Figure \ref{fig:LinearPeriods}. b) Fits to the A’(Li) versus $T_{\rm eff}$ relation of each cluster for G and K dwarfs.  Symbols have the same meaning as in Figures \ref{fig:ALi_teff} and \ref{fig:ALi_shift}. c) The predicted standard pre-MS Li depletion (at 100 Myr) for the metallicities of our clusters, interpolated from the models of P97. d) The difference A’(Li) - pre-MS Li depletion, $\Delta$Li; that is, fits from panel b minus panel c. e) For each of M48 and Hyades/Praesepe, stars with known P are separated into two groups: stars below the fit in Figure 8b that rotate (slightly) faster (light blue) and one with stars above the fit that rotate (slightly) slower. For the Pleiades, we first separate out the UFRs (large symbols) and then separate the remaining stars into faster and slower rotators. Note that to further illustrate the high Li in UFRs, we have added binary Pleiades UFRs to this panel, since binarity is unlikely (in this case) to affect the conclusion that UFRs have high Li.  Stars with unknown or uncertain P retain the default cluster colors (M48 is orange; Hyades/Praesepe is black).  For $T_{\rm eff}$ $<$ 5300 K, faster rotators in M48 and the Pleiades appear to have higher A’(Li) than slower rotators, on average.  No such trend is obvious for $T_{\rm eff}$ $>$ 5300 K.}
	\label{fig:comb_fig9}
\end{figure*}

Figure \ref{fig:LinearPeriods} shows rotational periods for the four clusters under consideration, and clearly illustrates the spin-down of stars as a function of age and $T_{\rm eff}$. Furthermore, the Period-$T_{\rm eff}$ relations for the three older clusters have less scatter than the relation for the Pleiades. The Pleiades does show a more populous group of slower rotators with relatively little scatter but there are also a number of UFRs. The fits for M48 and Hyades/Praesepe are for all stars, but for the Pleiades the fit is restricted to the more populous group of slower rotators. The fits stop at 4500 K because for the Pleiades and for M48 the typical scatter increases dramatically at lower $T_{\rm eff}$. Figure \ref{fig:comb_fig9} (panel a) shows the differences between these fits, $\Delta$P, to help investigate how J-loss might be related to Li depletion.

Before making this comparison we must take into account any (standard) pre-MS Li depletion at the base of the SCZ, which depends strongly on metallicity (P97; we ignore the possible complication that for $T_{\rm eff}$ $<$ $\sim$ 5000 K, additional Li depletion at the base of the SCZ may occur after 100 Myr). First, Figure \ref{fig:comb_fig9} (panel b) shows fits to the A’(Li)-$T_{\rm eff}$ relations for our clusters. For M48 and Hyades/Praesepe the fits go through the mean trends, but for the Pleiades the fit goes through the more populous group of lower A’(Li); below we will show that these tend to be slower rotators. Second, and keeping in mind the various uncertainties discussed by \citet{2014ApJ...790...72S} about the absolute standard depletion of Li, Figure \ref{fig:comb_fig9} (panel c) shows the predicted standard pre-MS Li depletion (at 100 Myr) for the metallicities of our clusters, interpolated from the models of P97.  Figure \ref{fig:comb_fig9} (panel d) shows the difference A’(Li) - pre-MS Li depletion, $\Delta$Li, which might be related to Li depletion due to J-loss.

We can now compare $\Delta$P to $\Delta$Li.  For the Pleiades and M48, $\Delta$P is fairly constant in going from 6000 K to 5300 K, and then decreases with lower $T_{\rm eff}$ (panel a). Since mass decreases with lower $T_{\rm eff}$, we might thus also expect stellar J content and perhaps J-loss to decrease from 6000 K to 5300 K and further still below 5300 K. In striking contrast to this trend, $\Delta$Li {\it increases} with decreasing $T_{\rm eff}$ throughout this $T_{\rm eff}$ range (panel d). This strongly suggests the importance of at least one additional parameter. We suspect the key is how the Li preservation region divides into the SCZ and the radiative region below where Li remains preserved. The depth of the SCZ increases with decreasing $T_{\rm eff}$, and the radiative Li preservation region correspondingly decreases (\citealt{1990ApJS...73...21D, 1990ApJS...74..501P}). Thus, to achieve a certain amount of $\Delta$Li, less J-loss is needed with decreasing $T_{\rm eff}$. The combined result is that greater Li depletion due to J-loss occurs for decreasing $T_{\rm eff}$ in spite of the diminishing J-loss with decreasing $T_{\rm eff}$.  We call upon new models to investigate these ideas.

For M48 and Hyades/Praesepe, $\Delta$P is fairly constant in going from 6000 K to 5500 K, and then {\it increases} with lower $T_{\rm eff}$, in contrast to M48 versus Pleiades.  For M48 versus Hyades/Praesepe, the $\Delta$Li follows the $\Delta$P more closely than in M48 versus Pleiades, where for M48 versus Hyades/Praesepe the $\Delta$Li is fairly constant from 6000 K to 5600 K and then increases. It is also interesting to note that $\Delta$P between M48 and Hyades/Praesepe is smaller than between M48 and Pleiades, and this is also true for $\Delta$Li.  So perhaps this also reinforces a connection between J-loss and Li depletion.

\begin{figure*}
	\centering
	\includegraphics[width=0.95\textwidth]{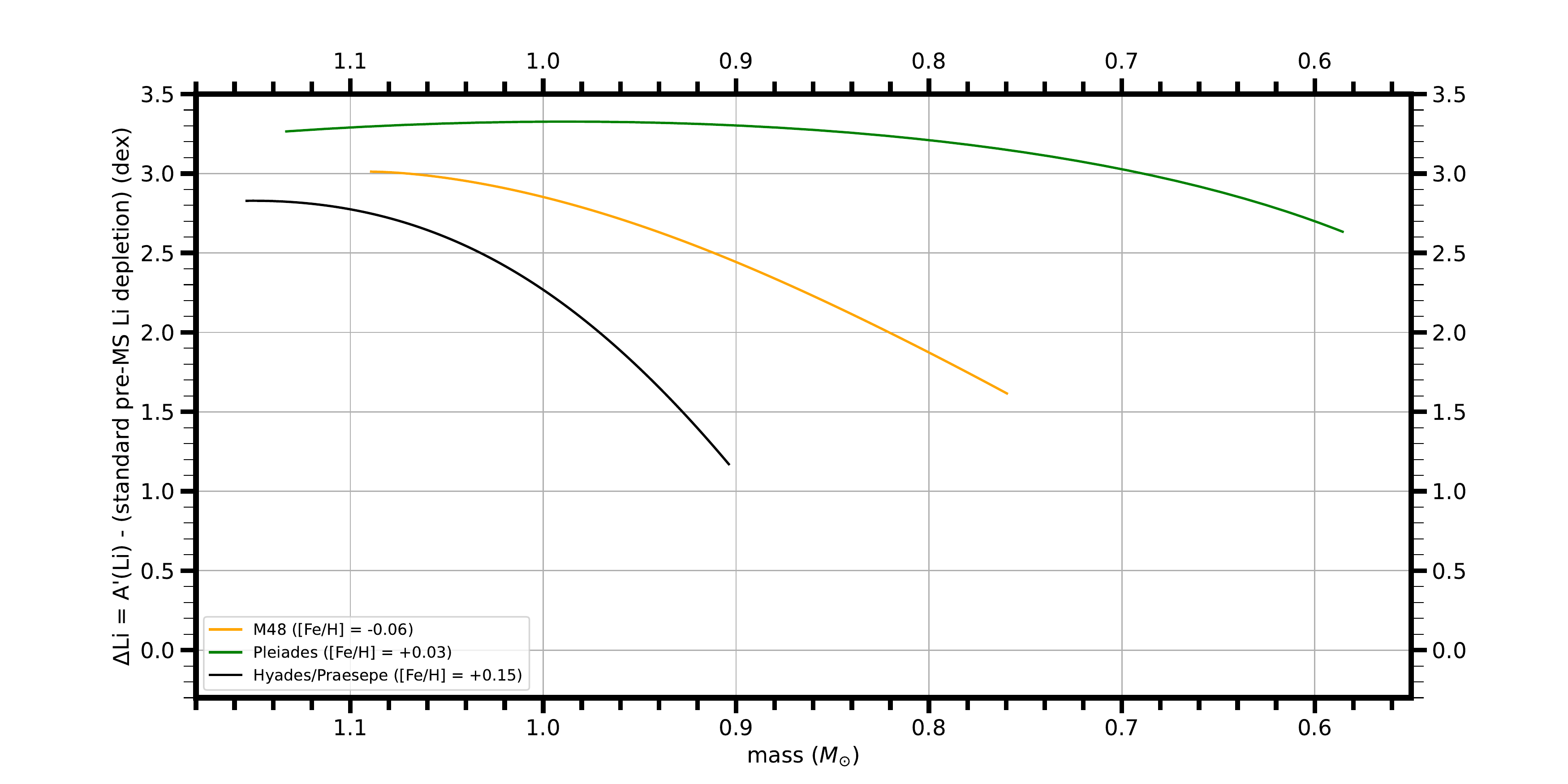}
	\caption{As with Figure \ref{fig:comb_fig9}, panel d, the difference A’(Li) - pre-MS Li depletion, $\Delta$Li, but this time plotted against stellar mass.}
	\label{fig:ali_depletion_mass}
\end{figure*}

Metallicity may also play a role and add complications.  At a given $T_{\rm eff}$, a star of higher metallicity has a higher mass.  Figure \ref{fig:ali_depletion_mass} shows $\Delta$Li as a function of mass.  Now $\Delta$Li increases with decreasing mass throughout the entire mass range shown. This could be consistent with the idea that the effectiveness of Li depletion induced by J-loss increases with decreasing mass, if the size of the radiative Li preservation region decreases with decreasing mass.  However, the size of this region might be more closely related to $T_{\rm eff}$ than to mass. Again, new models are needed to explore these ideas. Ideally, new models would take advantage of the constraints provided by both periods and Li depletion for all four clusters.

Next, we search for additional possible connections between rotational evolution and Li depletion.  Figure \ref{fig:comb_fig9} (panel e) shows the strikingly high A’(Li) in the UFRs, which has been noted earlier and will be discussed further below; here, we define a UFR as a star with P $<$ 1.2 d. We separate the remaining Pleiads into two groups: one below the fit of Figure 8b containing faster rotators and one above the fit containing slower rotators. As was also clearly shown in \citet{2018AA...613A..63B}, Figure \ref{fig:comb_fig9} shows that even when we exclude UFRs, for Pleiads with $T_{\rm eff}$ $<$ 5300 K, {\it faster rotators in this remaining group that excludes UFRs have higher Li abundances on average than slower rotators}.  How does this trend evolve?

To address this question, we similarly use periods to separate M48 and Hyades/Praesepe into two groups in each cluster of faster and slower rotators.  For $T_{\rm eff}$ $>$ 5300 K, neither M48 nor Hyades/Praesepe show any clear relation between period and A’(Li).  However, for M48 dwarfs with $T_{\rm eff}$ $<$ 5300 K, the faster rotators have higher A(Li), on average, than slower rotators.  Apparently the similar trend in the Pleiades survives through to the age of M48 of 420 Myr.  Unfortunately, at $T_{\rm eff}$ $<$ 5300 K the Hyades/Praesepe data remain silent on this issue as the A’(Li) vanish into upper limits.

Finally some remarks on the very interesting question of the Li evolution of the UFRs after the ages of the Pleiades and M35. \citet{2015MNRAS.449.4131S, 2015ApJ...807..174S} suggest that the high Li in Pleiades UFRs (and by inference M35 UFRs) can be explained by a combination of radius inflation (created by magnetic effects) and rotational stellar evolution.  In the absence of models discussing evolution beyond the Pleiades, we can only speculate about some possibilities, yet also possibly provide some interesting constraints.

On the one hand, Figure \ref{fig:LinearPeriods} shows that Pleiades late-G to early-K (5500 - 4500 K) UFRs have periods (at a given $T_{\rm eff}$) that are only 10-30\% of those of the majority of stars at the same -$T_{\rm eff}$; by contrast, the vast majority of stars in M48 with well-measured periods (as defined above in Section \ref{sec:rad}) in this $T_{\rm eff}$ range show variations in period of no more than $\sim$ 10-15\%.  So if M48 had UFRs in the past, they have lost more angular momentum than stars that were initially slower rotators, and might thus be expected to have rotationally-mixed and depleted more Li than slower rotators. But since such UFRs may have also had substantially higher A(Li) at the age of the Pleiades, it is not clear whether this increased depletion results in their ending up with Li above, near, or below the main trend at the age of M48.  Furthermore, this effect may be tempered by the fact that the Pleiades models for radius-inflated UFRs have shallower SCZs (which partly accounts for their higher A(Li)), so there is a larger (radiative) region below the SCZ that needs to be mixed by rotational mixing for Li depletion to occur.

Although Figure \ref{fig:comb_fig9} (panel e) shows that faster rotators in M48 with $T_{\rm eff}$ $<$ 5300 K have higher A’(Li) than slower rotators, we cannot know which stars in M48 might have been UFRs at the age of the Pleiades.  It is possible, perhaps even reasonable that the faster rotators in M48 correspond to an earlier UFR state, but we cannot be sure.  It might be helpful to observe more clusters with ages intermediate to those of the Pleiades and M48, to determine more explicitly and clearly how the Li abundances in UFRs evolve as the huge range in Pleiades periods shrinks to the much narrower range in M48 for 5400 K $>$ $T_{\rm eff}$ $>$ 4600 K.

\section{Summary } \label{sec:summary}

In Paper 1 (\citealt{2020AJ....159..220S}) we considered WIYN/Hydra spectra of 287 photometrically selected candidate members of the 420 Myr-old M48.  Here we consider WIYN/Hydra additional spectra of 70 stars, of which 42 were not observed in Paper 1.  Using similar methods as in Paper 1 to derive radial and rotational velocities, multiplicity and membership, we classify the 42 as follows:  18 are single members, 13 are binary members, 9 are members of uncertain multiplicity, 1 is a likely member of uncertain multiplicity, and 1 is a single likely non-member.  After combining with the stars from Paper 1, we have 234 cluster members and likely members of M48. 

We derive A(Li) for these 234 stars. For those stars that have detectable Li line, we use the {\it synth} task in MOOG to create synthetic spectra in the Li region. For candidates that have no detectable Li, we derive 3$\sigma$ upper limit Li abundances, which corresponds to the 3$\sigma$ upper limit equivalent widths computed from SNR, FWHM, and pixel scale.

The A(Li) - $T_{\rm eff}$ pattern of the M48 dwarfs (and 3 post-MS stars) is very informative. We compare to the 120 Myr-old Pleiades and the 650 Myr-old Hyades/Praesepe clusters, and discuss various features of the Li-$T_{\rm eff}$ relation and its evolution, divided into the following suggestive regions: Giants, A dwarfs (8600 - 7700 K), late A dwarfs (7700 - 7200 K), early F dwafs (7200 - 6650 K), the Wall near $T_{\rm eff}$ = 6700 K, the Li Dip (6675 - 6200 K), the Li Plateau (6200 - 6000 K), G dwarfs (6000 - 5150 K), and K dwarfs (5150 - 4000 K).

The three giants could be well-explained by the post-MS Li evolution tracks from \citet{2010A&A...522A..10C}, which include subgiant Li dilution and some earlier Li depletion during the MS due to rotational mixing.  Almost all A dwarfs have Li upper limits near or above the presumed initial cluster Li abundance, which is consistent with no Li depletion in these stars but is also consistent with an undetermined amount of Li depletion. Two of five late A dwarfs are clearly Li-rich.  Possible causes include diffusion, planetesimal accretion, and engulfment of a planet or Li-preserving brown dwarf; since some of these scenarios have different signatures, future data may be able to distinguish between them.

Our M48 data add to the large cadre of evidence supporting rotational mixing due to angular momentum loss as the dominant Li depletion mechanism in a wide range of spectral types, namely F, G, and K dwarfs. In particular, differences in Li at a given $T_{\rm eff}$ separates mechanisms than can create such differences (rotational mixing) from those than cannot (diffusion, gravity waves).  By way of example, we show comparison of spectra of seven pairs of dwarfs from early F dwarfs, the Li Dip, the Li Plateau, and G and K dwarfs. Each pair has the same $T_{\rm eff}$ but clearly different Li.

Early F dwarfs in older clusters show correlated depletion of Li with stellar spindown (\citealt{2019AJ....158..163D}). M48 shows evidence that the Li depletion begins at least as early as 420 Myr. 

The Li-$T_{\rm eff}$ trends of the Li Dip, Li Plateau, and G and K dwarfs are very clearly delineated and are intermediate to those of the Pleiades and the Hyades/Praesepe, which illustrates the Li depletion as a function of age. The cool side of the Li Dip is especially well-defined with little scatter. The Li-$T_{\rm eff}$ trend is very tight in the Li Plateau and early G dwarfs but scatter increases gradually for cooler dwarfs. We discuss how diffusion (near the Wall and slightly hotter) and gravity-wave driven mixing (in G and K dwarfs) may also play roles.

Ultra-Fast Rotators in Pleiades late-G/K dwarfs exhibit large Li overabundances. When the remaining stars are split into two groups, one with faster rotators and one with slower ones, the faster rotators have higher A(Li), on average than slower rotators. This trend appears to survive through to the age of M48.

Using periods in all four clusters, we discuss possible connections of angular momentum loss to Li depletion, including the possible roles of, a) increasing depth of the SCZ (with lower $T_{\rm eff}$) and the corresponding decrease in the size of the radiative Li preservation region below the SCZ, and b) metallicity.

Explaining the large Li over-abundances in late-G and K UFR dwarfs of clusters with ages near that of the Pleiades seems to require the additional effects of magnetic fields (which are themselves related to rotation) in creating radius inflation in those stars, together with rotational mixing (\citealt{2021MNRAS.500.1158J}). Our large M48 sample has no rapid rotators; if it ever had any they have all spun down. To explore how the A(Li) in rapid rotators evolves may require observations in clusters with ages intermediate to those of M48 and the Pleiades or M35, and perhaps closer to M35. Given the information now available on periods and Li abundances in all four clusters, we call on new models to explore the relationship between angular momentum loss and Li depletion and the possible connection to the size of the radiative Li preservation region below the SCZ, and the role of (standard) pre-MS Li depletion and its dependence on metallicity.

\begin{acknowledgements}
	
NSF support for this project was provided to C.P.D. through grant AST-1909456. Q.S. thanks support from the Shuimu Tsinghua Scholar Program. We also thank the WIYN 3.5m staff for helping us obtain excellent spectra.

This work has made use of data from the European Space Agency (ESA) mission Gaia, processed by the Gaia Data Processing and Analysis Consortium (DPAC). Funding for the DPAC has been provided by national institutions, in particular the institutions participating in the Gaia Multilateral Agreement.
	
\end{acknowledgements}

\bibliography{sun22}{}
\bibliographystyle{aasjournal}



\end{document}